**Multistep magnetization switching in orthogonally twisted ferromagnetic monolayers.**

Carla Boix-Constant[1], Sarah Jenkins[2], Ricardo Rama-Eiroa[3,2], Elton J. G. Santos[2,4,3]\*, Samuel Mañas-Valero[1,5]\*, Eugenio Coronado.[1]\*

[1] Instituto de Ciencia Molecular (ICMol) - Universitat de València, Catedrático José Beltrán 2, Paterna 46980, Spain.

[2] Institute for Condensed Matter Physics and Complex Systems, School of Physics and Astronomy, The University of Edinburgh, EH9 3FD, United Kingdom.

[3] Donostia International Physics Center (DIPC), 20018 Donostia-San Sebastián, Basque Country, Spain.

[4] Higgs Centre for Theoretical Physics, The University of Edinburgh, EH9 3FD, United Kingdom.

[5] Kavli Institute of Nanoscience, Delft University of Technology, Lorentzweg 1, Delft, 2628CJ, The Netherlands.

e-mail: samuel.manas@uv.es, esantos@ed.ac.uk, eugenio.coronado@uv.es

**The advent of twist-engineering in two-dimensional (2D) crystals enables the design of van der Waals (vdW) heterostructures exhibiting emergent properties. In the case of magnets, this approach can afford artificial antiferromagnets with tailored spin arrangements. Here, we fabricate an orthogonally-twisted bilayer by twisting 90 degrees two CrSBr ferromagnetic monolayers with an easy-axis in-plane anisotropy. The magneto-transport properties reveal multistep magnetization switching with a magnetic hysteresis opening, that is absent in the pristine case. By tuning the magnetic field, we modulate the remanent state and coercivity and select between hysteretic and non-hysteretic magneto-resistance scenarios. This complexity pinpoints spin anisotropy as a key aspect in twisted magnetic superlattices. Our results highlight the control over the magnetic properties in vdW heterostructures, leading to a variety of field-induced phenomena and opening a fruitful playground for creating desired magnetic symmetries and manipulating non-collinear magnetic configurations.**

Metamagnets and their field-induced phase transitions offer a plethora of counterintuitive phenomenology, as already quoted by Kramers,[1] with a direct competition between magnetic anisotropy, exchange, and dipolar energies.[2] In absence of magnetic field, these materials show zero net magnetization that suddenly increases until its saturation –thus, resembling a ferromagnet– above a certain magnetic field threshold.[1] A good example of an A-type metamagnet is offered by the layered vdW semiconductor CrSBr. The spins in every single layer (*ab* plane) couple ferromagnetically between them ($T_C$ ~150 K), pointing along the easy *b* axis, whereas the layers couple between them antiferromagnetically ($T_N$ ~ 140 K).[3] By applying a magnetic field, it is possible to flip the layers' magnetization in a parallel fashion via a spin reversal and to induce a spin reorientation along the magnetic field direction. This transition does not present hysteresis.[4–12] In bulk, the saturation fields at 2 K are 0.6 T, 1 T and 2 T for the easy (*b*), intermediate (*a*) and hard (*c*) magnetic axis, respectively.[10] This vdW material can be thinned down to the monolayer limit and integrated into electronic nano-devices. Upon the field-induced spin switching, the magneto-resistance (MR) is large and negative from bulk down to the bilayer case, with a reduction of the saturation field along the easy-axis (from 0.6 T in bulk to 0.2 T in the bilayer at 2 K).[8–10,13,14] The monolayer limit is characterized by the absence of MR for fields applied along the easy axis and small and positive MR for fields applied along the intermediate and hard axis.[10,13]

The ability for isolating, manipulating and twisting 2D crystals adds a new degree of control in vdW heterostructures, affording emergent new properties, like superconductivity in twisted bilayer graphene.[15] As far as the 2D magnetic materials are concerned, twisting is much less explored. Still, it has allowed the creation of new magnetic ground states. For example, by twisting small angles the 2D magnet $CrI_3$, a modulation of the spin-reversal by magneto-optical techniques has been reported.[16–18] This twist engineering not only produces a Moiré superlattice but can also induce Moiré magnetic exchange interactions, in which unique spin textures like magnetic skyrmion have been theoretically predicted.[19–22] However, no 2D twisted-magnets have been incorporated into electronic devices so far, remaining the magneto-transport effects in twisted-magnets fully unexplored.

Here, we twist by *ca.* 90 degrees two CrSBr ferromagnetic monolayers, thus forming an orthogonally-twisted bilayer. In analogy with the artificial antiferromagnets reported in synthetic spintronics –where the magnetic properties are tailored by growing multilayers of different antiferromagnets, in contrast with crystalline bulk antiferromagnets–,[23] this twisted heterostructure can be envisaged as an artificial antiferromagnetic bilayer. For probing its magneto-transport properties, this bilayer is integrated in a vertical vdW heterostructure formed by either few-layers graphene or metallic $NbSe_2$ thin-layers (**Fig. 1a-b**; see **Methods**).[24–27] Note that, in stark contrast with $CrI_3$, where the spins are out-of-plane, in CrSBr the spins are in-plane pointing along the easy magnetic *b*-axis, with an intermediate *a*-axis –also in-plane– but with a hard magnetic *c*-axis –out-of-plane direction–. This orthogonal configuration yields to an intriguing spin scenario where several terms might compete with an applied magnetic field as the Zeeman split energy, the inter-layer magnetic interactions (which favors an antiparallel orientation between the layers) and the local spin anisotropy at each CrSBr layer (which are perpendicular at the twisted configuration). This case is different from the common Moiré patterns in twisted bilayers, where a modification of the band structure is reached by twisting small angles.[15]

An example of an orthogonally-twisted-CrSBr heterostructure is shown in **Fig. 1a-b**. In this vertical geometry, the MR can be rationalized within a spin-valve picture, considering a two-current channel model: when the magnetization of both layers is antiparallel (parallel), there is a higher (lower) resistance across the heterostructure.[10,28,29] The field-dependence of the MR at 10 K is presented in **Fig. 1c** for in-plane magnetic fields aligned along the easy-axis of one of the layers (in this case, the top layer; α = β = 0º). Starting at high negative fields (red curve in **Fig. 1c**), the MR is negative and field independent down to -1 T; then, it increases until a maximum positive MR is observed at *ca.* +0.16 T. Above this field, it decreases again until reaching a saturation value above +1 T. This value coincides with that observed for the spin reorientation along the intermediate magnetic axis, *a*, thus suggesting that this is determined by the spin anisotropy. Reversing the magnetic field yields to a symmetrical curve that exhibits the maximum in MR at *ca.* -0.16 T (blue curve in **Fig. 1c**). These two curves cross at zero field (ZF) showing a hysteretic behavior when the field modulus is kept below *ca.* 0.32 T. For an easier visualization of the hysteresis, we present as a top panel in **Fig. 1c** the increment value, defined as $\Delta X = X_{+B \to -B} - X_{-B \to +B}$, where X states either for the resistance (R) or the MR while decreasing (+B→-B; blue curve in **Fig. 1c**) or increasing (-B→+B; red curve in **Fig. 1c**) the external magnetic field (B). Then, non-zero ΔX values indicate a hysteretic effect. As well, a zoom of the hysteretic region is presented in **Fig. 1d**, showing several resistance drops and plateaus and two lower limiting MR branches (with positive —red— and negative —blue— slopes) crossing at ZF. No relevant influence of the field sweeping rate is observed (**Supplementary Figure 1**). For a better comparative with the orthogonally-twisted bilayer, we show the corresponding MR behavior for pristine monolayer and bilayer CrSBr in **Fig. 1e-g**.[10] In the pristine case, the spin reversal takes place via a spin-flip for fields applied along the easy-magnetic axis and a spin-canting process for fields along the intermediate- and hard-magnetic axis.[7,8,10,13]

A qualitative understanding of the MR behavior of the orthogonally-twisted bilayer upon the application of a magnetic field along the easy- (intermediate-) magnetic axis of the top (bottom) monolayer (**Fig. 1c**) is as follows: at high negative fields (region from -3 T to -1 T) the magnetization of both layers is parallel ($\varphi = 0°$, where $\varphi$ is the angle formed between the magnetization of the top and bottom layer) and yielding to a state of low resistance according with a spin-valve picture. Below -1T the anisotropy is able to progressively reorient the magnetization of the bottom layer from its intermediate magnetic axis towards its easy-magnetic axis, while that of the top layer stays unchanged since the field is applied along its easy magnetic axis. As a consequence, from -1 T to 0 T an increase of the resistance is observed in agreement with the progressive increase of $\varphi$. In fact, at zero-field, the magnetization of both layers would be orthogonal ($\varphi = 90°$), assuming negligible inter-layer interactions. Upon the application of positive fields, the magnetization of the bottom layer continues the canting process ($\varphi > 90°$), tending to adopt an antiparallel configuration to satisfy the antiferromagnetic coupling, thus increasing the resistance to a maximum value at 0.16 T. At this point, the top layer flips its magnetization to be oriented along the positive magnetic field and $\varphi$ decreases ($\varphi < 90°$), thus yielding to a big drop of the resistance. Further magnetic fields tend to continue canting the magnetization of the bottom layer, thus decreasing $\varphi$ and, therefore, the resistance. Above 1 T (range from 1 T to 3 T), the magnetization of the top and bottom layers is parallel ($\varphi = 0°$) and the lower resistance state is observed. Decreasing magnetic fields yields to a symmetric configuration but observing the MR peak at negative fields and, consequently, yielding to a hysteretic effect (a detailed view of the process is presented in the **Supplementary Figure 2**). This scenario, which is possible thanks to the in-plane magnetic anisotropy of CrSBr, cannot be observed in twisted $CrI_3$, since it exhibits an out-of-plane anisotropy. This behavior is in sharp contrast with that of the pristine bilayer, which shows a single maximum of MR at ZF, as a result of the antiparallel orientation between the two layers, and no hysteretic effects (**Fig. 1e-g**). [10,13] Finally, we consider the in-plane angular dependence (**Supplementary Figure 3**). All the curves exhibit the same general trend discussed above but with different coercivity fields and $\Delta X$ values. The field orientation hence allows for a fine tuning and control of the hysteretic parameters. Note the asymmetry between 0º and 90º, suggesting that the underlying spin dynamics are dominated by one of the layers —as discussed later, it is due to the larger stray fields at the twisted layers— . We note that, for fields applied along directions different that the easy-magnetic axis, the reversal mechanism can be more complex since both layers can be canting, thus motivating future magnetic imaging experiments in these CrSBr twisted layers. Regarding magnetic fields applied along the hard-magnetic axis *c* (out-of-plane direction), a hysteretic behavior is manifested as well, but with a significantly broader maximum of MR (**Supplementary Figure 4**). In this case, the MR curves are saturating for fields above 2 T, which, as for the in-plane case, coincides with the field needed to reorient the spins along the magnetic field direction (*c* in the present case; **Fig. 1e-g**). Similar results are observed in different orthogonally-twisted bilayer CrSBr heterostructures, underlying the robustness of the observed phenomenology (**Supplementary Figure 5**), although the exact switching magnetic values differ between the different devices, probably due to slightly different twisting angles.

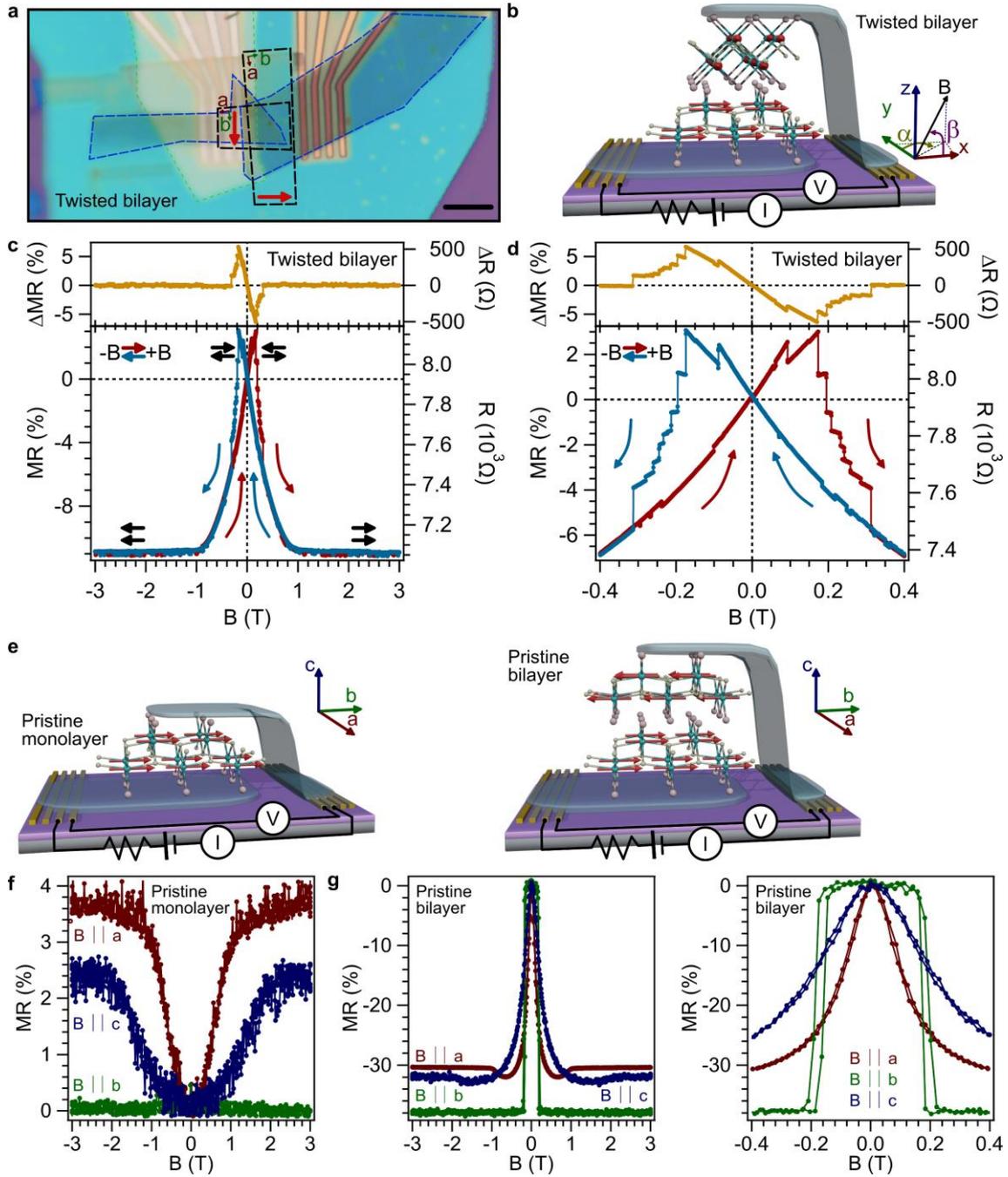

**Figure 1.- Magnetic field dependence of the magneto-resistance (MR) in orthogonally-twisted bilayer CrSBr. a,** Optical image of a vertical van der Waals heterostructure consisting of twisted CrSBr monolayers (black dashed lines) in between few-layers graphene (blue dashed lines). Different insulating h-BN layers (green dashed lines) are employed both for avoiding shortcuts and protecting the heterostructure. Red arrows indicate the easy-magnetic axis (*b*) of every CrSBr monolayer, being the intermediate-magnetic axis (*a*) perpendicular to it. The hard-magnetic axis (*c*) corresponds to the out-of-plane direction. Scale bar: 5 µm. **b,** Schematic view of the heterostructure (not to scale), highlighting the twisted CrSBr monolayers (pink, yellow and cyan balls correspond to bromine, sulfur and chromium atoms, respectively; red arrows represent the spin lying along the easy-magnetic axis, assuming negligible inter-layer magnetic interactions) placed in between few-layers graphene or NbSe$_2$ thin layers (blue color) on top of pre-patterned electrodes (gold color) together with a sketch of the electrical measurement configuration. **c-d,** Field-dependence of the resistance and MR (bottom panel) as well as its increment (top panel), defined as $\Delta X = X_{+B \to -B} - X_{-B \to +B}$, where X states either for the resistance or the MR at T = 10 K and $\theta = \varphi = 0º$. Sweeping up (down) trace is depicted in red (blue). Red/blue arrows indicate the sweeping direction of the magnetic field. Black arrows sketch the relative configuration of both layers' magnetization. MR is defined as MR (%) = 100·[R(B) − R(0)]/R(0). **e-g,** Reference experiments on pristine monolayer and bilayer based on our previous work,[10] including the corresponding sketches (panel **e**) and field dependence of the MR for fields applied along the easy (*b*), intermediate (*a*) and hard (*c*) magnetic axes for pristine monolayer (panel **f**) and bilayer (panel **g**) at T = 10 K.

Next, we consider both the field and temperature dependence of the MR (**Fig. 2**). We observe that the behavior resembles that reported for the pristine bilayer.[10] Upon cooling down the system, a negative MR starts developing below 200 K due to the onset of short-range interactions within the layers. Then MR reaches a broad plateau at *ca.* 150 K, near $T_c$, and below 100 K it increases again (**Fig. 2a**). However, some differences with the pristine bilayer are observed. First, in the pristine bilayer a minimum in MR, instead of a plateau, is observed at 150 K, followed at 100 K by a decrease. Second, a hysteretic behavior is observed from temperatures below $T_N$ (**Fig. 2b-d**), increasing the coercive field and ΔMR upon cooling down, while no hysteresis is observed in the pristine bilayer. Similar trends are observed for fields applied along different directions (**Supplementary Figure 6**).

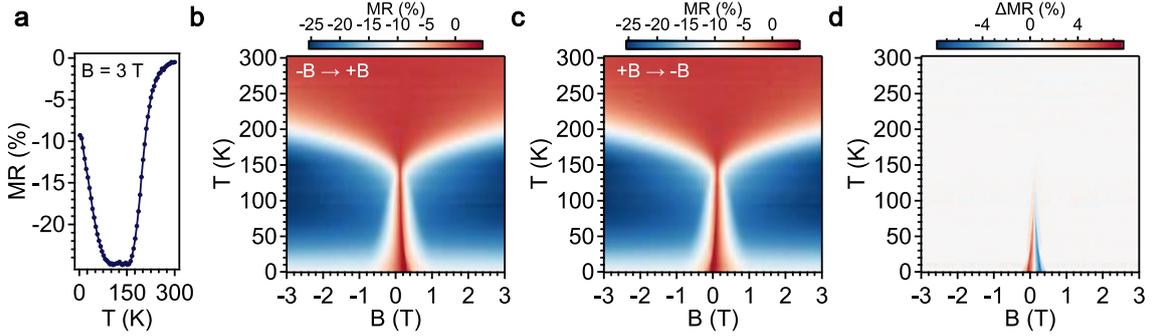

**Figure 2.- Field and temperature dependence of the MR in orthogonally-twisted bilayer CrSBr. a,** Temperature dependence of the MR at saturated fields (B = 3 T). **b-c,** Field and temperature dependence of the MR while sweeping from negative (positive) to positive (negative) fields. **d,** Field and temperature dependence of ΔMR. MR is defined as MR (%) = 100·[R(B) – R(0)]/R(0), being R(0) the resistance obtained at zero field and ΔMR= MR$_{+B\rightarrow-B}$ – MR$_{-B\rightarrow+B}$. θ = φ = 0º.

To further explore the irreversibility of the observed hysteresis in **Fig. 1**, we perform a series of first-order reversal curves (FORC). The FORC analysis lies behind the Preisach model.[30,31] We increment sequentially the maximum applied magnetic field ($B_{max}$) in steps of 20 mT, after an initial saturation at negative fields (sequence: -0.6 T → +$B_{max}$ → -0.6 T). Selected curves are shown in **Fig. 3a** (see the **Supplementary Video 1** for the whole data set). For sweeping fields below *ca.* 0.1 T (|$B_{max}$| = 0.06 T in **Fig. 3a**), the resistance increases/decreases upon increasing/decreasing B following the behavior already observed in **Fig. 1.d** when sweeping from negative fields (limiting branch with positive slope). No hysteresis is observed for this loop, being the MR curve symmetric (ΔMR = 0 at ZF). A more interesting scenario is offered when this field threshold is overcome (|$B_{max}$| = 0.16 T in **Fig. 3a**). In this case, the resistance increases (red curve) upon increasing B, as before, until a sharp drop occurs at *ca.* 0.1 T. Then, upon decreasing B (blue curve), the resistance decreases but with a smaller slope until a second drop is observed at *ca.* -0.1 T, when it returns to the initial path (limiting branch with positive slope). This behavior results in the emergence of an asymmetric hysteresis (ΔMR ≠ 0 at ZF). Similar asymmetric curves with successive drops in the resistance, giving rise to steps and plateaus at well-defined magnetic fields, are observed upon increasing the maximum sweeping magnetic field value (|$B_{max}$| = 0.18 T in **Fig. 3**, **Supplementary Video 1** and **Supplementary Figure 7**). Interestingly, each step observed for positive fields is characterized by a different slope while returning to ZF. This slope decreases until a saturation field is reached (0.32 T in the present case). For B > 0.32 T, the limiting branch with negative slope is reached and the hysteresis loop becomes fully symmetric with respect to the R axis (|$B_{max}$| = 0.50 T in **Fig. 3a**). Interestingly, when coming from positive saturated fields (sequence: +0.6 T → -$B_{max}$ → +0.6 T in **Fig. 3b** and **Supplementary Video 1**), the same phenomenology is observed but reversing the modulus of the switching fields (mirror image with respect to the R axis).

Therefore, this magnetic system is formed by two ground states that are degenerated at ZF but that evolve with opposite MR slopes in presence of B. Thus, an initial saturation at negative fields leads to a state defined by the MR branch with positive slope (**Fig. 3a**). This state is sketched as a set of blue circles in the **Fig. 3**. Conversely, when coming from positive fields, a different state is obtained (set of red circles in **Fig. 3**) leading to the MR branch with negative slope. For B values within the range ±0.32 T the system evolves hysteretically and selectively towards one of these two ground states and only for higher |B| values a change of ground state is possible. This allows to select at will the ground state of the system. Furthermore, in the hysteretic region such evolution takes place through successive steps at specific fields that may be associated to intermediate states. This multistep phenomenology can be related with the Preisach model. Thus, starting from one of the two MR branches, each one of the resistance drops observed in the hysteresis curves is associated to the switch of an individual hysteron, leading to each one of the intermediate states postulated above. In applied terms, every one of these switches could be potentially employed as a bit of information. This is schematically sketched in **Fig. 3** by sweeping red/blue bytes. Importantly, there is also magnetic memory at ZF since we can select between hysteretic and non-hysteretic MR scenarios depending on the initial ground state of the system. In the **Supplementary Figure 8**, we consider different magnetic field sweep protocols and, for example, in the sequence ZF → +$B_{max}$ → ZF we observe hysteresis only after an initial saturation in negative magnetic fields. Therefore, the magnetic history allows us to control the appearance or not of hysteresis. To manifest the robustness of these results, we present in **Supplementary Figure 9** the study for other orthogonally-twisted CrSBr bilayers. Overall, similar trends, although at different switching fields, are observed under different in-plane field orientations (**Supplementary Figure 10**) and temperatures (**Supplementary Figure 11**).

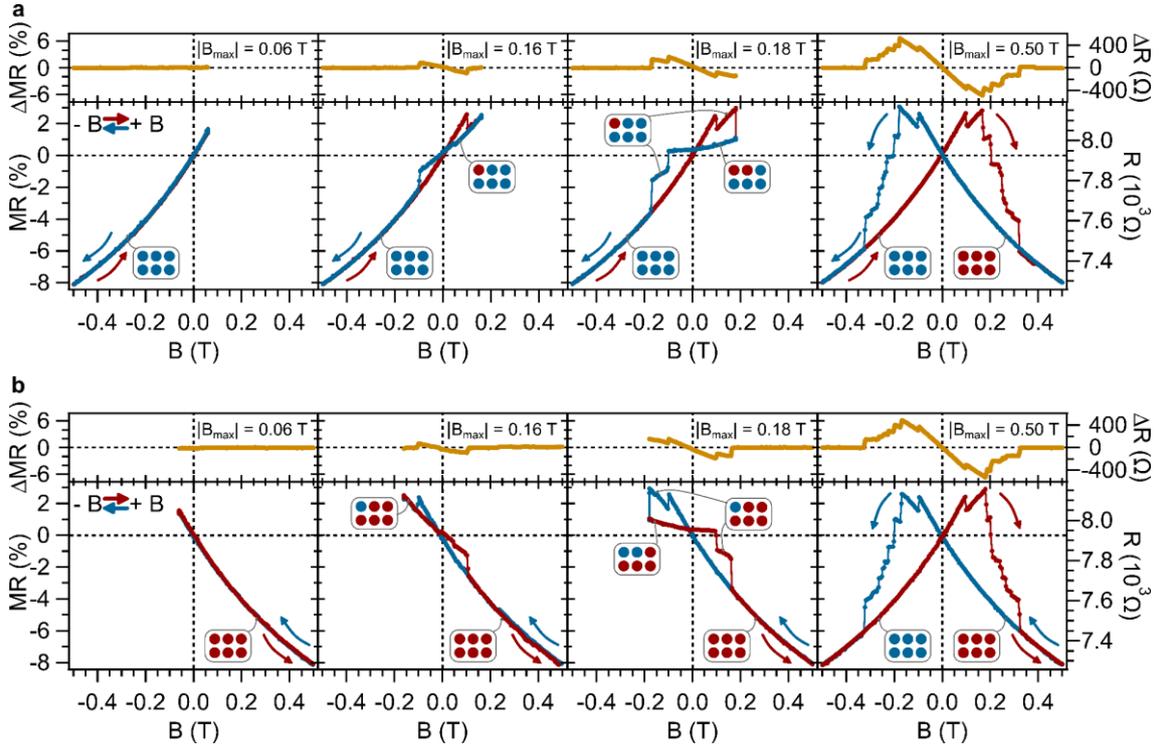

**Figure 3.- Multistep magnetization switching with magnetic memory in orthogonally-twisted bilayer CrSBr.** First-order reversal curves considering the sequence a) -0.6 T → +$B_{max}$ → -0.6 T and b) +0.6 T → -$B_{max}$ → +0.6 T at 10 K and θ = φ = 0 °. $B_{max}$ is incremented sequentially in steps of 20 mT and selected curves are shown (see **Supplementary Video** for the whole dataset). The saturated state at negative (positive) magnetic fields is schematically sketched as a set of blue (red) circles configuration, being every spin switch related to the change of one individual hysteron (squared hysteresis operator characterized by a coercive field and a field shift from zero) within the Preisach model. MR is defined as MR (%) = 100·[R(B) − R(0)]/R(0), being R(0) the resistance obtained at zero field in the symmetric case.

Regarding the origin of the multistep magnetization switching, we can attribute it to the stabilization of different domain configurations and spin textures as unveiled by atomistic spin dynamic simulations. We have considered the case of a CrSBr-based orthogonally twisted bilayer, where the top monolayer is rotated 90 degrees with respect to the bottom one (inset in **Fig. 4a**). The size of the simulation system is 100 nm × 100 nm along *x*- and *y*-axes, with no periodic boundary conditions along the aforementioned directions, and a cell thickness along the *z*-axis corresponding to two-unit cells to accommodate the two stacked monolayers (see **Methods** for details). In line with the experimentally-based measurement protocol, we apply a simulated field of varying strengths along the x-axis, *i.e.*, along the easy (intermediate) magnetic axis of the bottom (top) layer. Note that, for an easier visualization of the results, the easy (intermediate) magnetic axis of the bottom (top) layer is rotated if compared with **Fig. 1a-d**. We then simulate the field-cooling from 200 K (above $T_N$) to 0 K using spin dynamics techniques for 2D magnets[32–34]. In this way, it is possible to follow microscopically the variation of the magnetic features at the final simulated state of the system for different field strengths. We evaluate the angle $\theta$ between the magnetic moment vector $\mathbf{M}/M_s$ (where $M_s$ is the volumetric saturation magnetization) of the top monolayer and the *x* direction, as it provides a strong descriptor of the spin orientations at the layers. Interestingly, we have observed that when low fields are applied $(0.01 - 0.095\text{T})$, the magnetization of the top monolayer is canted from its easy *b*-axis towards its intermediate *a*-axis (**Fig. 4a,b**). If we increase the magnitude of the magnetic field to $0.10 - 0.14$ T (**Fig. 4a,c**), we observe the appearance of non-collinear spin configurations in the form of hybrid domain walls (Bloch-type) in the top monolayer. Intriguingly, this type of magnetic configurations, for this range of field strengths, only occurs if chiral spin-interactions like Dzyaloshinskii-Moriya interactions (DMI) are considered in the simulations,[35] causing a magnetic frustration due to the competing contributions. This could lead to the appearance of more complex non-collinear spin distributions for larger systems. When the applied field is increased further to $0.18 - 0.30$ T (**Fig. 4a,d**), the Zeeman-like contribution will overpower the internal fields causing the magnetization of the top monolayer to align along the magnetic field direction, that is, along its intermediate magnetic axis, *a*. The illustration of each spin phase at specific field magnitudes is displayed in **Fig. 4e-g** with the snapshots extracted from the simulations in **Supplementary Figure 12**. **Supplementary Movies S2-S4** show the entire evolution of the dynamics at 0.06 T, 0.10 T and 0.20 T, respectively. For instance, we observed that at 0.10 T (**Fig. 4e** and **Supplementary Movie S3**) the domain wall profile of the top layer flips from the +*y* to -*y* direction and the spins at the centre are along the applied field direction, *x*, parallel to the bottom monolayer. We observed that these different spin textures are not present on the pristine bilayer CrSBr as expected, since both layers have their easy-axes along the same direction. Moreover, we have applied temperature to the system (5 K) and the simulated results remain consistent despite of the thermal fluctuations and noise (**Supplementary Figure 13**). It is worth mentioning that as one of the layers is twisted (*e.g.,* top layer), the dipolar fields $H_{dip}$ generated at that layer become larger relative to the untwisted layer (*e.g.,* bottom layer), which conditionate the response of the system (**Supplementary Figure 14**). That is, one of the layers becomes more dominant than the other inducing the appearance of some of the MR effects discussed before in the measurements. Indeed, the variations of stray-fields with the applied field follow those observed in the spin textures with the formation of canting fields and domain walls (**Supplementary Figure 12**). This suggests that the dynamic evolution of the magnetisation with the external magnetic field follows a Barkhausen-like effect trend[36] with a series of sudden changes in the size and orientation of the magnetic domains. In our case, however, since the top layer is twisted with respect to the underlying layer, a systematic flip of the spins with the field is possible until saturation is reached. The smaller fields to saturate the system in the simulations (~0.2 T) relative to the measurements (~ 1 T) may be due to variations of the magnetic parameters used,[37,38] but the overall picture is well described and in sound agreement with the measurements.

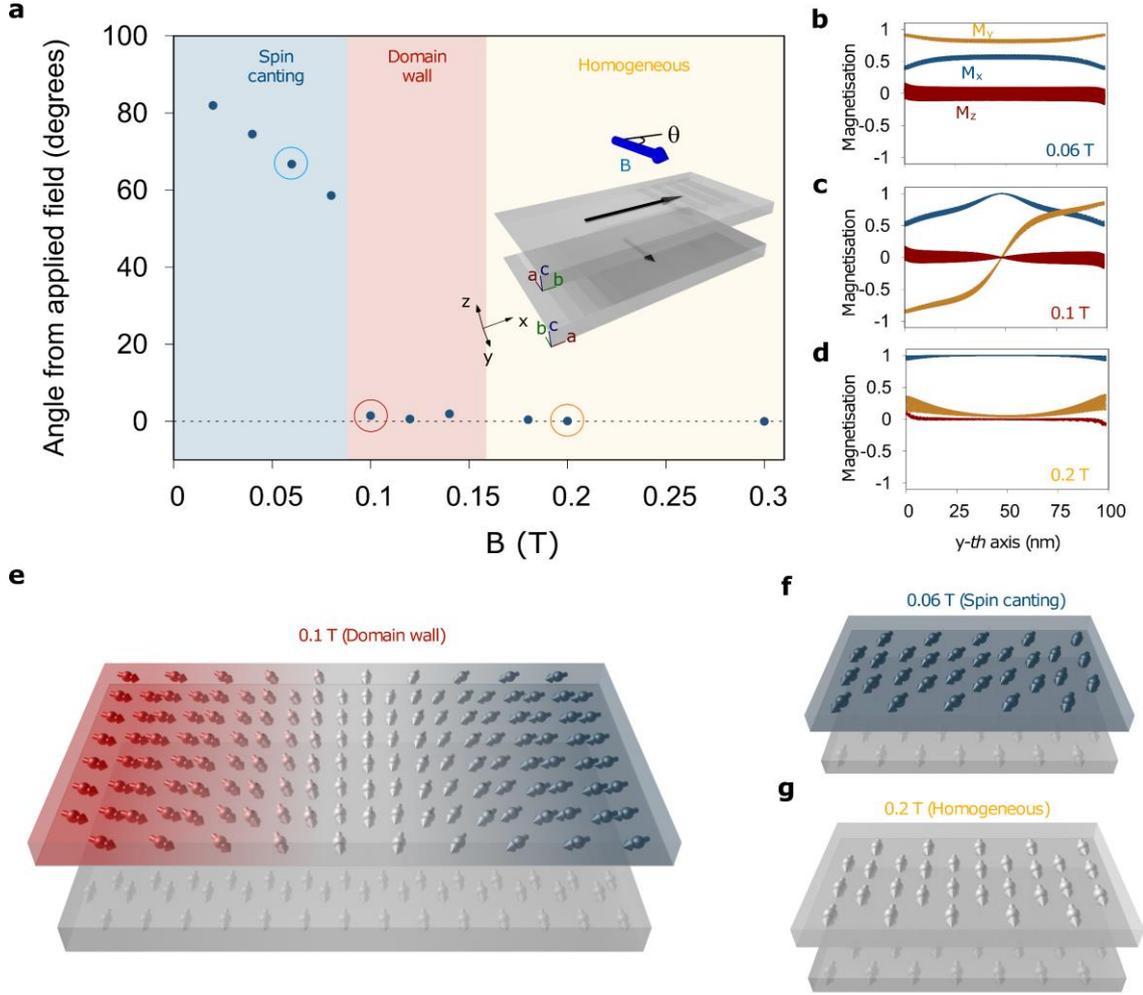

**Figure 4.- Field-induced spin-textures in orthogonally-twisted bilayer CrSBr. a,** After cooling under different applied field strengths (0.01 – 0.3 T) applied along $x$ –parallel (perpendicular) to easy (intermediate) magnetic axis of the bottom (top) layer–, we calculate the angle $\theta$ between the magnetisation direction (**M/Ms**) of the top layer to the applied field **B** (blue arrow). The field is initially applied along one of the easy-axis of the layers (dark arrows), and as its magnitude increases **M/Ms** changes accordingly to be aligned with **B** (see inset). Three magnetic phases can be stabilised with the applied field: spin canting (the spins are aligned but at an angle between the anisotropy easy-axis direction of the top layer and the applied field direction), domain wall (part of the spins of the top layer orient along of the field, and another with the bottom layer underneath which induced the formation of domain walls) and homogeneous (both layers have their spins aligned with the field). Three values of the field (0.06 T, 0.1 T, 0.2 T) are highlighted with circles and further analysed in the following panels as an example. The crystallographic a-, b- and c-axes for every monolayer are indicated. **b-d,** Projections of the magnetisation $M_x$, $M_y$ and $M_z$ at 0 K as a function of the position (nm) along the $a$-axis of the top layer at 0.06 T, 0.1 T and 0.2 T, respectively. **e-g,** Schematics of the spin configuration observed in the spin dynamics simulations (**Supplementary Figure 12**) at 0.1 T (domain wall), 0.06 T (spin canting) and 0.2 T (homogeneous).

In conclusion, we have shown that twisting engineering of magnetic 2D materials is a fruitful platform for the emergence of new correlated phases in artificial metamagnets, as exemplified here by the appearance of multistep spin switching accompanied by hysteretic MR effects in orthogonally-twisted bilayer CrSBr. These field-induced features can be controlled by playing with the modulus and direction of the applied magnetic field, being absent in pristine CrSBr mono- and bi-layers. Overall, our results pinpoint twisted bilayer CrSBr as a versatile and rich platform for controlling and addressing the magnetic information on 2D magnets —of special relevance in areas such as spintronics or magnonics[39]—, as well as for motivating a new playground for fundamental studies. In particular, this orthogonally-twisted bilayer CrSBr may offer a promising route for the creation and manipulation of non-colinear magnetic textures, like vortices or topologically protected skyrmions and merons.[20,21] On the other hand, the controlled stacking of 2D magnetic monolayers under defined angles opens new avenues to increase the magnetic symmetry in the plane, thereby reducing the anisotropy energy. Of special interest is to reach the crossover from easy-axis to easy-plane anisotropy, since easy-plane (XY) systems[40] are predicted to host dissipationless spin transport.[41,42]

**Acknowledgements**

The authors acknowledge the financial support from the European Union (ERC AdG Mol-2D 788222, FET OPEN SINFONIA 964396), the Spanish MCIN (2D-HETEROS PID2020-117152RB-100, co-financed by FEDER, and Excellence Unit "María de Maeztu" CEX2019-000919-M) and the Generalitat Valenciana (PROMETEO Program, PO FEDER Program IDIFEDER/2018/061, a Ph.D fellowship to C.B.-C., and APOSTD-CIAPOS2021/215 to S.M.-V). This study forms part of the Advanced Materials program and was supported by MCIN with funding from European Union NextGenerationEU (PRTR-C17.I1) and by Generalitat Valenciana. E.J.G.S. acknowledges computational resources through CIRRUS Tier-2 HPC Service (ec131 Cirrus Project) at EPCC (http://www.cirrus.ac.uk) funded by the University of Edinburgh and EPSRC (EP/P020267/1); ARCHER2 UK National Supercomputing Service via Project d429. E.J.G.S. acknowledges the Spanish Ministry of Science's grant program ``Europa-Excelencia'' under grant number EUR2020-112238, the EPSRC Open Fellowship (EP/T021578/1), and the Edinburgh-Rice Strategic Collaboration Awards for funding support. S.M.-V. acknowledges the support from the European Commission for a Marie Skłodowska–Curie individual fellowship No. 101103355 - SPIN-2D-LIGHT. We thank Á. López-Muñoz for his constant technical support and fundamental insights, A. Bedoya-Pinto for helpful discussions as well as C. Boudeau for the isolation and assembly of CrSBr monolayers.

**Author Contributions Statement**

C.B.-C. fabricated the orthogonally-twisted bilayers and performed the magneto-resistance measurements, with the participation of S.M.-V. S.M.-V. and C.B.-C. performed the crystal growth and characterization of CrSBr crystals, performed the data analysis and wrote the first version of the manuscript, under the supervision of E.C. S.J. and R.R.-E. performed atomistic spin dynamic simulations under the supervision E.J.G.S. The work was conceived by S.M.-V.

**Competing Interests Statement**

The authors declare no competing interests.

**References**

1. Stryjewski, E. & Giordano, N. Metamagnetism. *Adv Phys* **26**, 487–650 (1977).

2. Hellwig, O., Kirk, T. L., Kortright, J. B., Berger, A. & Fullerton, E. E. A new phase diagram for layered antiferromagnetic films. *Nat Mater* **2**, 112–116 (2003).

3. Lee, K. *et al.* Magnetic Order and Symmetry in the 2D Semiconductor CrSBr. *Nano Lett* **21**, 3511–3517 (2021).

4. Göser, O., Paul, W. & Kahle, H. G. Magnetic properties of CrSBr. *J Magn Magn Mater* **92**, 129–136 (1990).

5. López-Paz, S. A. *et al.* Dynamic magnetic crossover at the origin of the hidden-order in van der Waals antiferromagnet CrSBr. *Nat Commun* **13**, 4745 (2022).

6. Lee, K. *et al.* Magnetic Order and Symmetry in the 2D Semiconductor CrSBr. *Nano Lett* **21**, 3511–3517 (2021).

7. Wilson, N. P. *et al.* Interlayer electronic coupling on demand in a 2D magnetic semiconductor. *Nat Mater* **20**, 1657–1662 (2021).

8. Telford, E. J. *et al.* Layered Antiferromagnetism Induces Large Negative Magnetoresistance in the van der Waals Semiconductor CrSBr. *Advanced Materials* **32**, 2003240 (2020).

9. Wu, F. *et al.* Quasi-1D Electronic Transport in a 2D Magnetic Semiconductor. *Advanced Materials* **34**, 2109759 (2022).

10. Boix-Constant, C. *et al.* Probing the Spin Dimensionality in Single-Layer CrSBr Van Der Waals Heterostructures by Magneto-Transport Measurements. *Advanced Materials* **34**, 2204940 (2022).

11. Zur, Y. *et al.* Magnetic Imaging and Domain Nucleation in CrSBr Down to the 2D Limit. *Advanced Materials* (2023) doi:10.1002/adma.202307195.

12. Marques-Moros, F., Boix-Constant, C., Mañas-Valero, S., Canet-Ferrer, J. & Coronado, E. Interplay between Optical Emission and Magnetism in the van der Waals Magnetic Semiconductor CrSBr in the Two-Dimensional Limit. *ACS Nano* **17**, 13224–13231 (2023).

13. Telford, E. J. *et al.* Coupling between magnetic order and charge transport in a two-dimensional magnetic semiconductor. *Nat Mater* **21**, 754–760 (2022).

14. Ye, C. *et al.* Layer-Dependent Interlayer Antiferromagnetic Spin Reorientation in Air-Stable Semiconductor CrSBr. *ACS Nano* **16**, 11876–11883 (2022).

15. Cao, Y. *et al.* Unconventional superconductivity in magic-angle graphene superlattices. *Nature* **556**, 43–50 (2018).

16. Xie, H. *et al.* Twist engineering of the two-dimensional magnetism in double bilayer chromium triiodide homostructures. *Nat Phys* **18**, 30–36 (2022).

17. Song, T. *et al.* Direct visualization of magnetic domains and moiré magnetism in twisted 2D magnets. *Science (1979)* **374**, 1140–1144 (2021).

18. Xu, Y. *et al.* Coexisting ferromagnetic–antiferromagnetic state in twisted bilayer CrI3. *Nat Nanotechnol* **17**, 143–147 (2022).

19. Yang, B., Li, Y., Xiang, H., Lin, H. & Huang, B. Moiré magnetic exchange interactions in twisted magnets. *Nat Comput Sci* **3**, 314–320 (2023).

20. He, Z. *et al.* Multiple Topological Magnetism in van der Waals Heterostructure of $MnTe_2$/$ZrS_2$. *Nano Lett* **23**, 312–318 (2023).

21. Kim, K.-M., Kiem, D. H., Bednik, G., Han, M. J. & Park, M. J. Theory of Moiré Magnets and Topological Magnons: Applications to Twisted Bilayer CrI3. *arXiv 2206.05264* (2022).

22. Dang, W. *et al.* Electric-Field-Tunable Spin Polarization and Carrier-Transport Anisotropy in an A-Type Antiferromagnetic van der Waals Bilayer. *Phys Rev Appl* **18**, 064086 (2022).

23. Duine, R. A., Lee, K. J., Parkin, S. S. P. & Stiles, M. D. Synthetic antiferromagnetic spintronics. *Nature Physics* vol. 14 217–219 Preprint at https://doi.org/10.1038/s41567-018-0050-y (2018).

24. Boix-Constant, C. *et al.* Out-of-Plane Transport of $1T-TaS_2$/Graphene-Based van der Waals Heterostructures. *ACS Nano* **15**, 11898–11907 (2021).

25. Boix-Constant, C., Mañas-Valero, S., Córdoba, R. & Coronado, E. Van Der Waals Heterostructures Based on Atomically-Thin Superconductors. *Adv Electron Mater* **7**, 2000987 (2021).

26. Boix-Constant, C. *et al.* Strain Switching in van der Waals Heterostructures Triggered by a Spin-Crossover Metal–Organic Framework. *Advanced Materials* **34**, 2110027 (2022).

27. Torres, K. *et al.* Probing Defects and Spin-Phonon Coupling in CrSBr via Resonant Raman Scattering. *Adv Funct Mater* **33**, 2211366 (2023).

28. Huang, B. *et al.* Electrical control of 2D magnetism in bilayer $CrI_3$. *Nat Nanotechnol* **13**, 544–548 (2018).

29. Klein, D. R. *et al.* Probing magnetism in 2D van der Waals crystalline insulators via electron tunneling. *Science* **360**, 1218–1222 (2018).

30. Mayergoyz, I. D. Mathematical Models of Hysteresis. *Phys Rev Lett* **56**, 1518–1521 (1986).

31. Arzuza, L. C. C. *et al.* Domain wall propagation tuning in magnetic nanowires through geometric modulation. *J Magn Magn Mater* **432**, 309–317 (2017).

32. Jenkins, S. *et al.* Breaking through the Mermin-Wagner limit in 2D van der Waals magnets. *Nat Commun* **13**, 6917 (2022).

33. Wahab, D. A. *et al.* Quantum Rescaling, Domain Metastability, and Hybrid Domain-Walls in 2D $CrI_3$ Magnets. *Advanced Materials* **33**, 2004138 (2021).

34. Kartsev, A., Augustin, M., Evans, R. F. L., Novoselov, K. S. & Santos, E. J. G. Biquadratic exchange interactions in two-dimensional magnets. *NPJ Comput Mater* **6**, 150 (2020).

35. Casas, B. W. *et al.* Coexistence of Merons with Skyrmions in the Centrosymmetric Van Der Waals Ferromagnet $Fe_{5-x}GeTe_2$. *Advanced Materials* **35**, 2212087 (2023).

36. Tebble, R. S. The Barkhausen Effect. *Proceedings of the Physical Society. Section B* **68**, 1017–1032 (1955).

37. Bowden, G. J., Stenning, G. B. G. & van der Laan, G. Inter and intra macro-cell model for point dipole–dipole energy calculations. *Journal of Physics: Condensed Matter* **28**, 066001 (2016).

38. Bo, X., Li, F., Xu, X., Wan, X. & Pu, Y. Calculated magnetic exchange interactions in the van der Waals layered magnet CrSBr. *New J Phys* **25**, 013026 (2023).

39. Esteras, D. L., Rybakov, A., Ruiz, A. M. & Baldoví, J. J. Magnon Straintronics in the 2D van der Waals Ferromagnet CrSBr from First-Principles. *Nano Lett* **22**, 8771–8778 (2022).

40. Bedoya-Pinto, A. *et al.* Intrinsic 2D-XY ferromagnetism in a van der Waals monolayer. *Science (1979)* **374**, 616–620 (2021).

41. Takei, S. & Tserkovnyak, Y. Superfluid spin transport through easy-plane ferromagnetic insulators. *Phys Rev Lett* **112**, (2014).

42. König, J., Bønsager, M. C. & MacDonald, A. H. Dissipationless spin transport in thin film ferromagnets. *Phys Rev Lett* **87**, 187202-1-187202–4 (2001).

**Methods**

*Crystal growth:*

CrSBr crystals were synthesized by chemical vapor transport and characterized by powder and crystal X-Ray diffraction, energy dispersive X-Ray analysis, high-resolution transmission electron microscopy, SQUID magnetometry and temperature-dependent single crystal, as reported in our previous work.[10]

*van der Waals heterostructure fabrication:*

2D layers were obtained by mechanical exfoliation from their bulk counterparts under strict inert conditions (argon glovebox) since CrSBr monolayers degrade in air.[13,27] The obtained flakes were examined by optical microscopy (NIKON Eclipse LV-100 optical microscope under normal incidence) as a fast tool for identifying the number of layers and compared with our previously calibrated values.[10] Typical CrSBr flakes exhibit a ribbon shape, being the long (short) direction associated with the *a* (*b*) axis and being the *c* axis the out-of-plane direction, as verified by optical contrast, Raman spectroscopy and selected area electron diffraction patterns. Details are reported in our previous work.[10] The van der Waals heterostructures were fabricated by assembling the different layers by the deterministic assembly of the flakes using polycarbonate and with the help of a micromanipulator. Thus, the twisted-monolayers were placed between top and bottom few-layers metallic $NbSe_2$ or few-layers graphene, where several insulating h-BN layers were inserted both for avoiding possible short-cuts and protecting the whole heterostructure from degradation. The stack of 2D materials was placed on top of pre-lithographed electrodes (5 nm Ti/50 nm Au on 285 nm $SiO_2$/Si from NOVA Electronic Materials, LCC). The whole process was performed under inert atmosphere conditions.

A total of three orthogonally-twisted CrSBr bilayers were fabricated (device 1 -data shown in the main text- is based on metallic $NbSe_2$ thin-layers while device 2 and 3 -data shown as extended data- are based on few-layers graphene), observing a consistent phenomenology between all of them. Note that, in the case of using few-layers graphene, the intrinsic MR arising from the few-layers graphene is observed as well (in special, for out-of-plane applied magnetic fields), yielding to a finite positive value of the MR even at room temperature. Nonetheless, the magnetic fingerprints of the twisted-CrSBr are well noticeable, clearly developing below $T_N$.

In particular, device 1 is formed by a top (bottom) CrSBr monolayer of 77.2 µm² (53.3 µm²), with an overlap area of 9.3 µm² and a twisted angle of 92.5°. Device 2 is formed by a top (bottom) CrSBr monolayer of 190.1 µm² (117.3 µm²), with an overlap area of 15.9 µm² and a twisted angle of 89.3°. Device 3 is formed by a top (bottom) CrSBr monolayer of 206.6 µm² (121.1 µm²), with an overlap area of 7.9 µm² and a twisted angle of 87.0°.

*Magneto-transport measurements:*

Electrical measurements were performed in a Quantum Design PPMS-9 cryostat with a 4-probe geometry, where a DC current was passed by the outer leads and the DC voltage drop was measured in the inner ones. DC voltages and DC currents were measured (MFLI from Zurich Instruments) using an external resistance of 1 MΩ, *i.e.*, a resistance much larger than the sample. Temperature sweeps were performed at 1 K·min⁻¹, field sweeps at 200 Oe/s, rotation sweeps at 3 °/s and the current bias was 1 µA, unless otherwise explicitly specified. Magneto-resistance (MR) is defined as: MR = 100[R(B) – R(0)]/R(0), where B is the external magnetic field and R(0) is the resistance at zero field in the symmetric case (see text).

*Atomistic spin dynamic simulations:*

Our simulations were performed using atomistic spin dynamics simulation techniques [32–35,43–47] as implemented in the VAMPIRE software package.[47] The energetics of the system is described by the spin Hamiltonian:

$$\mathcal{H} = -\sum_{i<j} \mathbf{S}_i^\alpha J_{ij}^{\alpha\beta} \mathbf{S}_j^\beta + \sum_{i<j} \mathbf{D}_{ij} \cdot (\mathbf{S}_i \times \mathbf{S}_j) - k_a \sum_i (\mathbf{S}_i \cdot \hat{\mathbf{a}})^2 - k_b \sum_i (\mathbf{S}_i \cdot \hat{\mathbf{b}})^2$$

$$- \sum_i \mu_{s,i} \mathbf{S}_i \cdot \mathbf{B} + \mathcal{H}_\mathcal{D}$$

where $\mathbf{S}_i$ and $\mathbf{S}_j$ are unit vectors describing the local spin directions on Cr sites. The first term is the symmetric Heisenberg exchange and $J_{ij}$ is the exchange tensor between Cr sites, being $\alpha, \beta = a, b, c$. The first input is the Heisenberg exchange, which in CrSBr has seven intra-layer exchange terms ($J_{1-7}$) occurring between atoms within the same monolayer and two inter-layer terms ($J_{z1}$ and $J_{z2}$) occurring from one monolayer to another. The value of the second intra-layer nearest neighbour exchange ($J_2$) was taken from Wang, H. *et al.*[38] and was used as a reference to define the magnitude of the $J_{ij}$ elements, which are outlined below. In order to obtain satisfactory predictions of the critical temperature of CrSBr-based systems, the relative ratios between exchange parameters have been taken from Bo, X. *et al.*[48] For the inter-layer interactions, $J_{z1}$ and $J_{z2}$, we have used the values for the unrotated bilayer due to the absence of dramatic changes in the intermonolayer distances. The distances for the $J_{z1}$ only differ by about 5.44% and the average deviation in the $J_{z2}$ interactions is only 2.66%. As commented below, variations of these magnitudes do not change the results.

The second term is the anti-symmetric exchange or DMI which stabilizes topological states, where $\mathbf{D}_{ij}$ is the DMI vector. Due to the absence of inversion symmetry between interacting Cr-based atoms,[49] we have included the reported anti-symmetric contributions with DMI unit vectors parallel to the *a-th* (mediating $J_3$) and *b-th* (mediating $J_1$) axes, whose values are given, respectively, by $D_1 = 0.07$ meV and $D_3 = 0.18$ meV.[48]

The third term is the on-site anisotropy energy, which is made up of two uniaxial terms, where the relative values of the anisotropy constants, $k_a = 8.06$ meV and $k_b = 31.53$ meV, govern the intermediate *a-th* and easy *b-th* axes of the system.[38] It is important to note that previously introduced single-ion anisotropies are not, theoretically, the only ones that should contribute to the overall magnetocrystalline anisotropy of the system. The larger spin-orbit coupling of the Br atoms compared to those of Cr points to the existence of in-basal-plane-based exchange anisotropy terms at the previously defined spin Hamiltonian.[50] However, in the computational characterization of the system for the twisted bilayer we have chosen not to include them, despite the fact that it has been reported that they share the same order of magnitude as the on-site anisotropy contributions. This is because these second-ion terms can induce the *a-th* axis to be the easiest one in the system.[38] Moreover, the single-ion contributions are enough to unravel the main features observed experimentally. It is worth noting that, due to the rotation process, the easy-axis of the top monolayer is directed along the *a-th* spatial direction and the intermediate one the *b-th* axis (orthogonally directed with respect to the untwisted bottom monolayer).

The fifth term is the Zeeman energy, where **B** represents the externally applied magnetic field and $\mu_s$ is the atomic magnetic moment, to which the value $\mu_s = 2.88\mu_B$ has been assigned in consonance with the bulk scenario,[37] being $\mu_B$ the Bohr magneton.

The final term is the long-range dipole-dipole interaction, $\mathcal{H}_\mathcal{D}$, which can be expressed as:

$$\mathcal{H}_\mathcal{D} = \frac{\mu_0 \mu_s^2}{4\pi} \sum_j \left[ \frac{3\hat{\mathbf{r}}_{ij}(\hat{\mathbf{r}}_{ij} \cdot \mathbf{S}_j) - \mathbf{S}_j}{|\mathbf{r}_{ij}|^3} \right]$$

where $|\mathbf{r}_{ij}|$ is the distance between site i to j.

We also calculated the inter-layer exchange field as:

$$\mathbf{H}_{\text{exc}}^{\text{inter}} = \frac{1}{\mu_s} \left[ -4|J_{z1}|(\mathbf{m}_{\text{bottom}} + \mathbf{m}_{\text{top}}) + J_{z2}(m_{\text{bottom}}^c + m_{\text{top}}^c)\hat{\mathbf{c}} \right]$$

being the magnetisation of the bottom and top layer represented by $\mathbf{m}_{\text{bottom}}$ and $\mathbf{m}_{\text{bottom}}$, respectively. Taking into account that there are four nearest-neighbours, mediated by $J_{z1}$, and one next-nearest-neighbour, mediated by $J_{z2}$, interactions, we can estimate a maximum exchange field of ~0.15 T if we assume that $\mathbf{m}_{\text{bottom}}$ and $\mathbf{m}_{\text{top}}$ are fully parallel. This magnitude is much smaller than the dipolar fields induced by the twisted layer (**Supplementary Figure 14**), and suggests that variations of the order of 5-10% in exchange interactions will not affect the results in case the rotation might play a role. This correlates with potential variations due to the interlayer distance between Cr sites as commented above.

**Methods-only references:**

43. Augustin, M., Jenkins, S., Evans, R. F. L., Novoselov, K. S. & Santos, E. J. G. Properties and dynamics of meron topological spin textures in the two-dimensional magnet CrCl3. *Nat Commun* **12**, 185 (2021).

44. Abdul-Wahab, D. *et al.* Domain wall dynamics in two-dimensional van der Waals ferromagnets. *Appl Phys Rev* **8**, 041411 (2021).

45. Alliati, I. M., Evans, R. F. L., Novoselov, K. S. & Santos, E. J. G. Relativistic domain-wall dynamics in van der Waals antiferromagnet MnPS3. *NPJ Comput Mater* **8**, 3 (2022).

46. Wang, Q. H. *et al.* The Magnetic Genome of Two-Dimensional van der Waals Materials. *ACS Nano* **16**, 6960–7079 (2022).

47. Evans, R. F. L. *et al.* Atomistic spin model simulations of magnetic nanomaterials. *Journal of Physics: Condensed Matter* **26**, 103202 (2014).

48. Wang, H., Qi, J. & Qian, X. Electrically tunable high Curie temperature two-dimensional ferromagnetism in van der Waals layered crystals. *Appl Phys Lett* **117**, (2020).

49. Scheie, A. *et al.* Spin Waves and Magnetic Exchange Hamiltonian in CrSBr. *Advanced Science* **2202467**, 2202467 (2022).

50. Telford, E. J. *et al.* Designing Magnetic Properties in CrSBr through Hydrostatic Pressure and Ligand Substitution. *Advanced Physics Research* (2023) doi:10.1002/apxr.202300036.

**Multistep magnetization switching in orthogonally twisted ferromagnetic monolayers.**

Carla Boix-Constant[1], Sarah Jenkins[2], Ricardo Rama-Eiroa[3,2], Elton J. G. Santos[2,4,3]*, Samuel Mañas-Valero[1,5]*, Eugenio Coronado.[1]*

[1] Instituto de Ciencia Molecular (ICMol) - Universitat de València, Catedrático José Beltrán 2, Paterna 46980, Spain.

[2] Institute for Condensed Matter Physics and Complex Systems, School of Physics and Astronomy, The University of Edinburgh, EH9 3FD, United Kingdom.

[3] Donostia International Physics Center (DIPC), 20018 Donostia-San Sebastián, Basque Country, Spain.

[4] Higgs Centre for Theoretical Physics, The University of Edinburgh, EH9 3FD, United Kingdom.

[5] Kavli Institute of Nanoscience, Delft University of Technology, Lorentzweg 1, Delft, 2628CJ, The Netherlands.

e-mail: samuel.manas@uv.es, esantos@ed.ac.uk, eugenio.coronado@uv.es

**Table of contents**

# Section A - Supplementary Figures 1 - 16

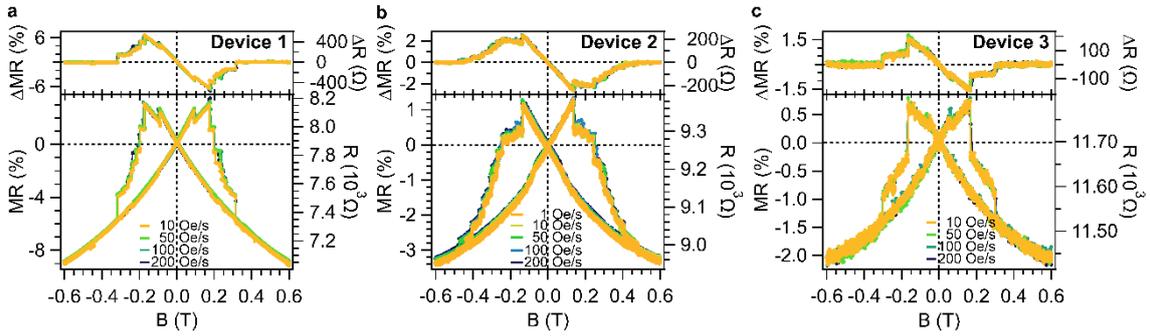

**Supplementary Figure 1.- Magnetic field dependence of the magneto-resistance (MR) in orthogonally-twisted bilayer CrSBr for different devices and field sweep rates. a,** Device based on $NbSe_2$ vertical van der Waals heterostructure (T = 10 K). **b-c,** Devices based on few-layers graphene vertical van der Waals heterostructures (T = 2 K). The field is applied in-plane ($\beta = 0°$) along the easy-axis of the CrSBr monolayer with smaller area, corresponding to $\alpha = 0°$ (for **a** and **b**) and $\alpha = 90°$ (for **c**). MR is defined as MR (%) = 100·[R(B) – R(0)]/R(0), being R(0) the resistance obtained at zero field.

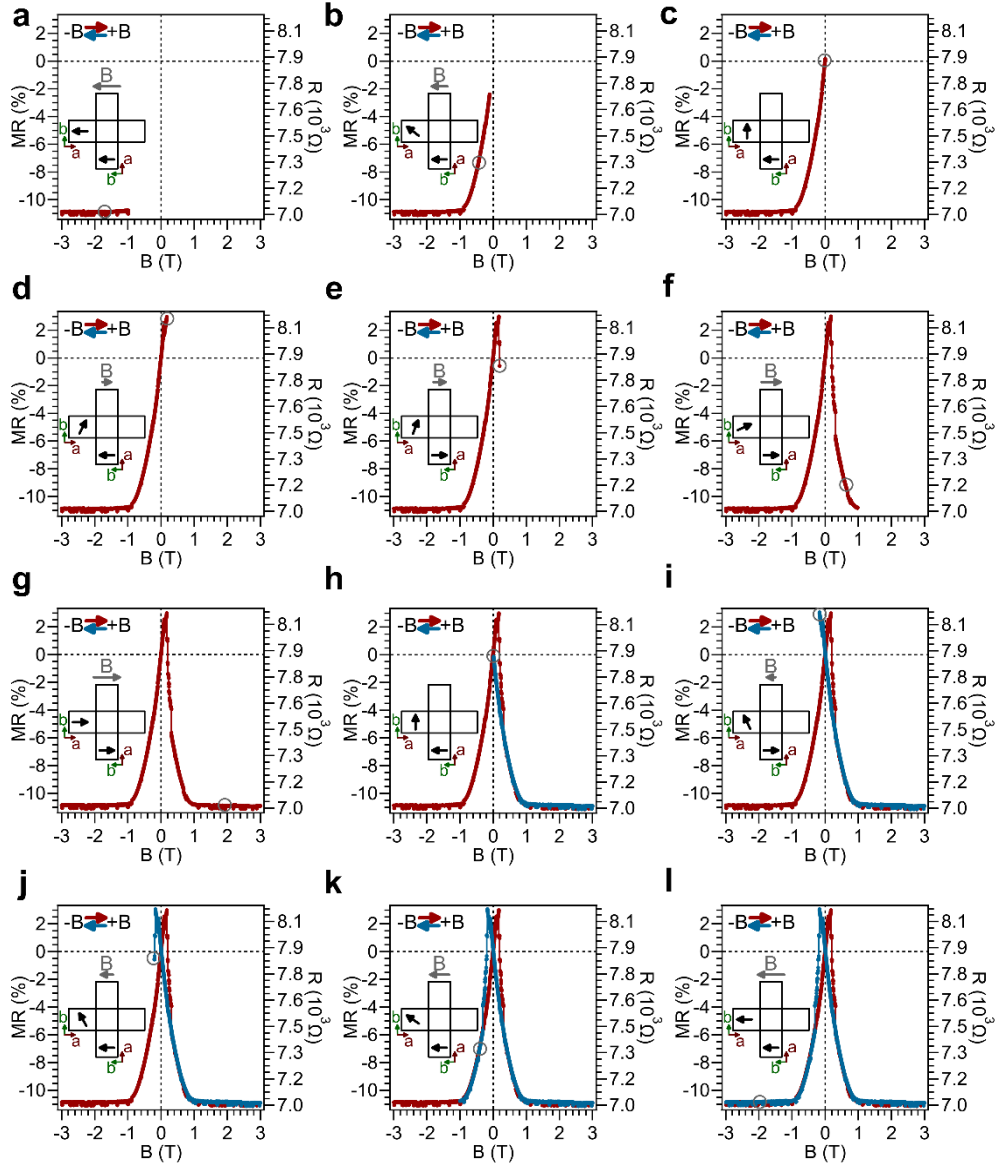

**Supplementary Figure 2.- Magnetization switching for an orthogonally-twisted bilayer CrSBr.** The applied external magnetic field (represented as grey arrow and position marked as grey circle) is aligned with the easy-axis (*b*) of one of the monolayers and aligned with the intermediate-axis (*a*) of the *rotated* monolayer. The magnetization for every layer is represented as a black arrow. **a,** At high negative magnetic fields, the magnetization of both layers is parallel and aligned with the field and, therefore, the resistance is minimum within a spin-valve model. **b,** Below -1 T, the magnetization of the *rotated* monolayer starts canting towards its easy-axis. The magnetization of both layers is not parallel, yielding to an increase of the resistance. **c,** At zero field, the magnetization of both layers is orthogonal assuming negligible inter-layer interactions. **d,** At small positive magnetic fields, the magnetization of the *rotated* monolayer starts canting towards the direction of the magnetic field. The angle between the magnetization of the two layers increases towards an antiparallel state and, therefore, the resistance increases. **e,** At *ca.* 0.16 T, the monolayer with its easy-axis along the field flips its magnetization (spin flip). Thus, the angle between the magnetization of both layers suddenly decreases and, then, the resistance drops. **f,** Applying higher positive magnetic fields cants the magnetization of the *rotated* monolayer towards the direction of the magnetic field. The angle between the magnetization of both layers decreases and, therefore, the resistance diminishes. **g,** For fields above 1 T, the magnetization of both layers is parallel along the direction of the applied field and the resistance is minimum. **h,** For fields below 1 T, the magnetization of the *rotated* layer starts canting towards its easy axis, yielding to a resistance increase. At zero field, the magnetization of both layers is orthogonal assuming negligible inter-layer interactions. **i,** At small negative magnetic fields, the magnetization of the *rotated* monolayer starts canting towards the direction of the magnetic field. The angle between the magnetization of the two layers increases towards an antiparallel state and, therefore, the resistance increases. **j,** At *ca.* -0.16 T, the monolayer with its easy-axis along the field flips its magnetization (spin flip). Thus, the angle between the magnetization of both layers suddenly decreases and, then, the resistance drops. **k,** Applying higher negative magnetic fields cants the magnetization of the *rotated* monolayer towards the direction of the magnetic field. The angle between the magnetization of both layers decreases and, therefore, the resistance diminishes. **l,** For fields below -1 T, the magnetization of both layers is parallel along the direction of the applied field and the resistance is minimum.

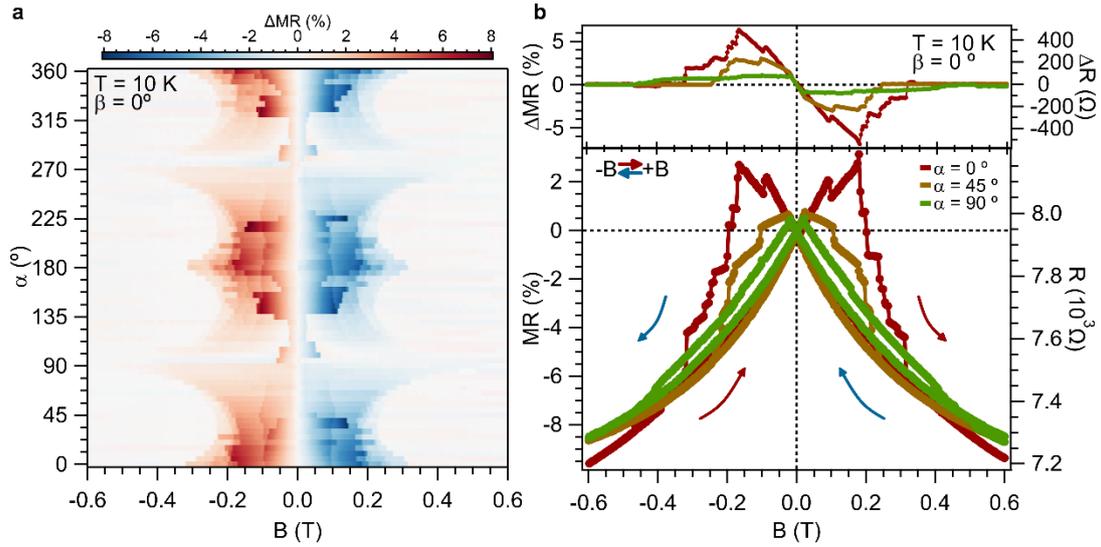

**Supplementary Figure 3.- In-plane magnetic field dependence of the magneto-resistance (MR) in orthogonally-twisted bilayer CrSBr (device 1). a,** 2D plot of ΔMR. **b,** Selected MR/resistance hysteresis loops (bottom panel) and its increment (top panel) at selected angles. Measurements corresponds to an orthogonally-twisted CrSBr bilayer based on metallic $NbSe_2$ thin-layers vertical van der Waals heterostructure. MR is defined as MR (%) = 100·[R(B) – R(0)]/R(0), being R(0) the resistance obtained at zero field.

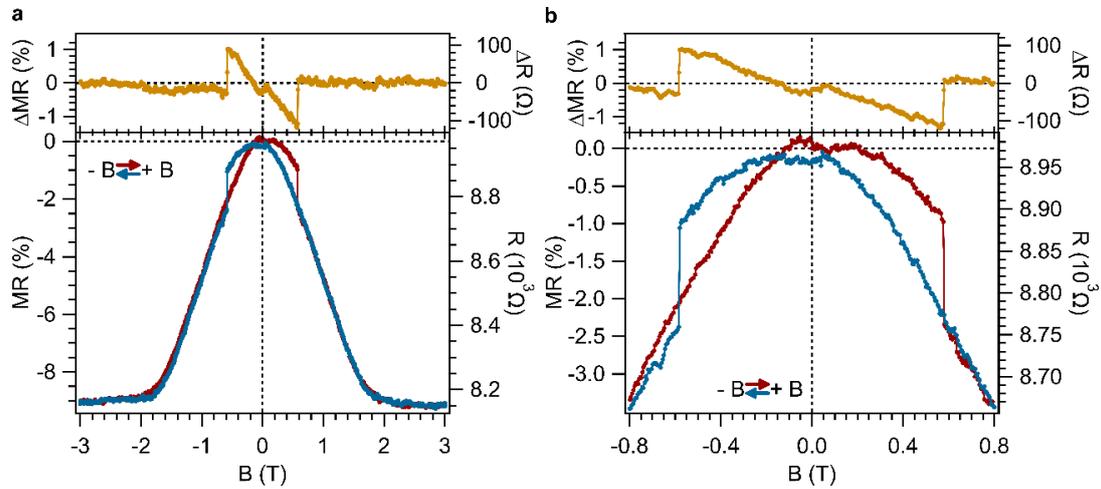

**Supplementary Figure 4.- Out-of-plane magnetic field dependence of the magneto-resistance (MR) in orthogonally-twisted bilayer CrSBr (device 1). a,b,** Field-dependence of the resistance and MR (bottom panel) as well as its increment (top panel), defined as $\Delta X = X_{+B\rightarrow-B} - X_{-B\rightarrow+B}$, where X states either for the resistance or the MR (T = 10 K, α = 0° and β = 90°). Sweeping up (down) trace is depicted in red (blue).

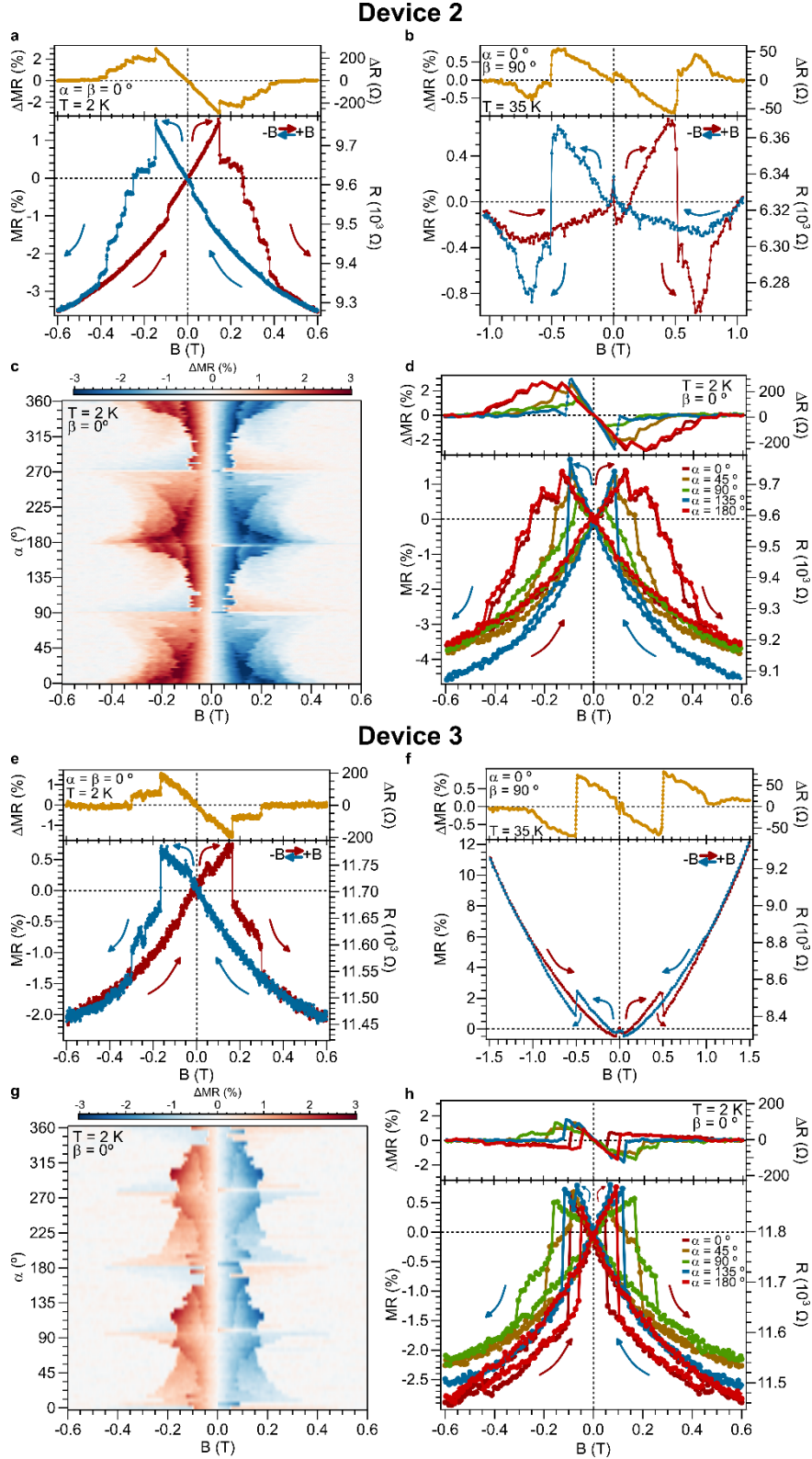

**Supplementary Figure 5.- Magnetic field dependence of the magneto-resistance (MR) in orthogonally-twisted bilayer CrSBr based on few-layers graphene van der Waals heterostructures.** Panels **a-d** (**e-h**) correspond to device 2 (3). **a,b,e,f,** Field-dependence of the resistance and MR (bottom panel) as well as its increment (top panel), defined as $\Delta X = X_{+B \to -B} - X_{-B \to +B}$, where X states either for the resistance or the MR for in-plane (**a,e** panels) and out-of-plane (**b,f** panels) fields. Sweeping up (down) trace is depicted in red (blue). Red/blue arrows indicate the sweeping direction of the magnetic field. MR is defined as MR (%) = 100·[R(B) – R(0)]/R(0). **c,g,** 2D plot of ΔMR. **d,h,** Selected MR and resistance hysteresis loops (bottom panel) and its increment (top panel) at selected angles. Note that the intrinsic MR arising from the few-layers graphene is observed as well (in special, for out-of-plane applied magnetic fields), yielding to a finite positive value of the MR even at room temperature. Nonetheless, the magnetic fingerprints of the twisted-CrSBr are well noticeable.

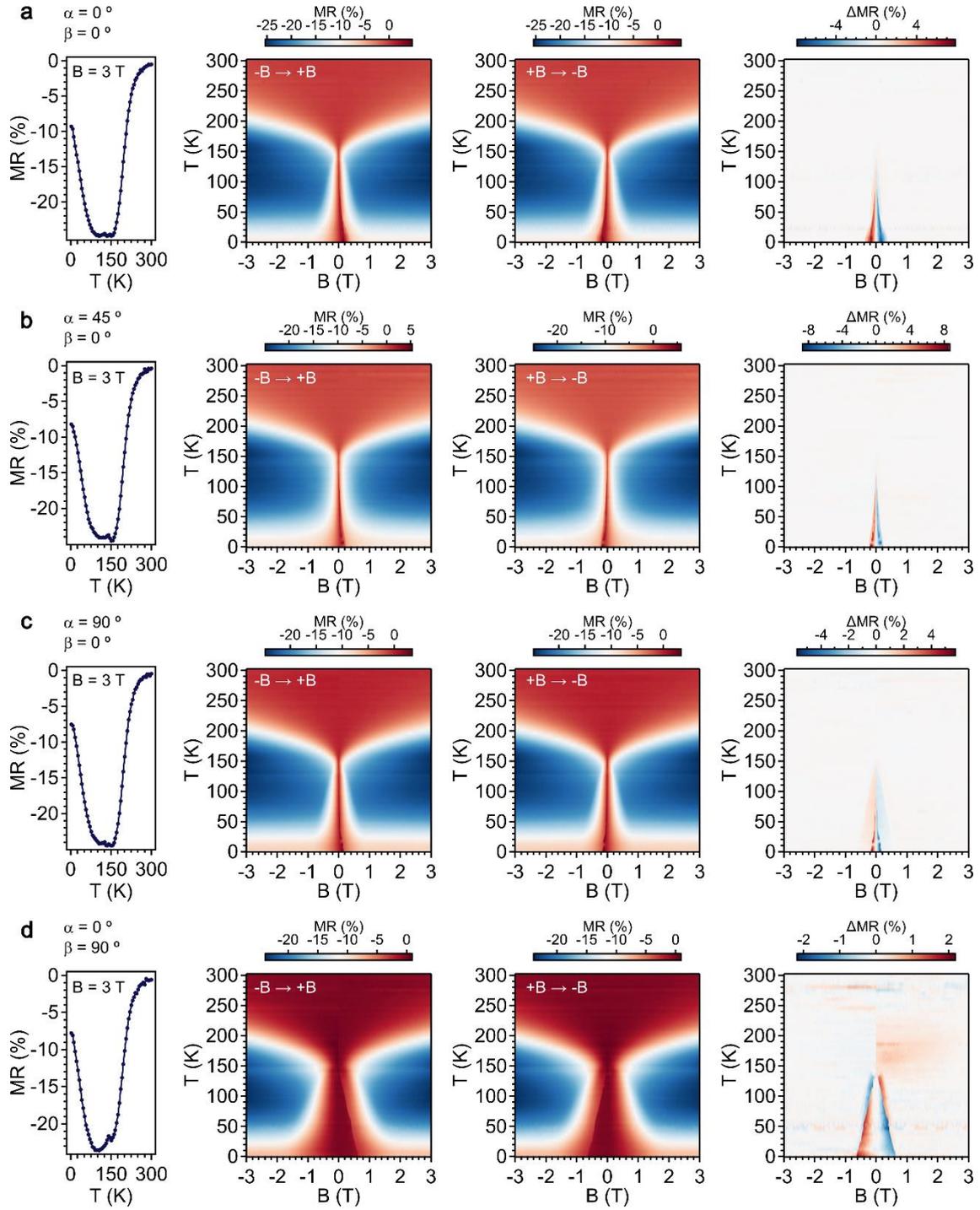

**Supplementary Figure 6.- Temperature and magnetic field dependence in orthogonally-twisted bilayer CrSBr. a-c,** In-plane field orientations. **d,** Out-of-plane orientation. First panel: Temperature dependence of the magneto-resistance (MR) in the saturated state (B = 3 T). Second (third) panel: field and temperature dependence of the MR while sweeping from negative (positive) to positive (negative) fields. Fourth panel: field and temperature dependence of ΔMR. MR is defined as MR (%) = 100·[R(B) – R(0)]/R(0), being R(0) the resistance obtained at zero field.

S6

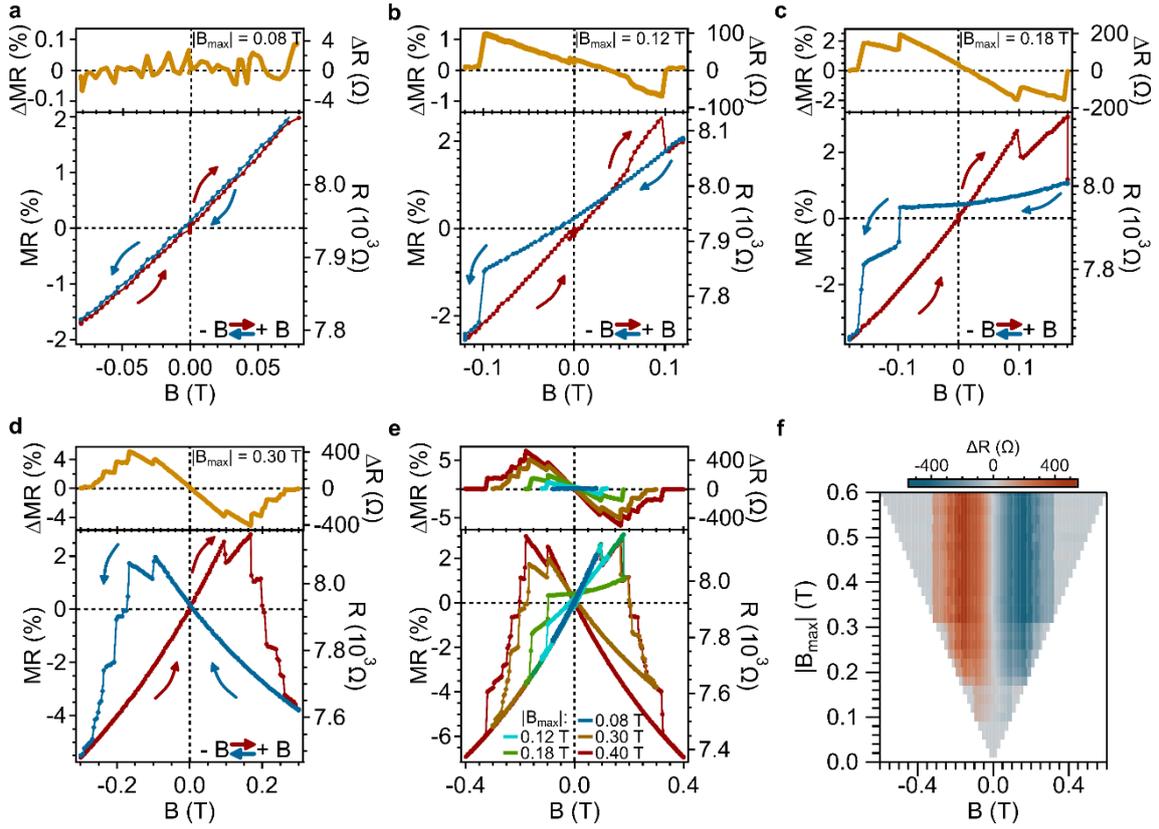

**Supplementary Figure 7.- Hysteresis opening in orthogonally-twisted bilayer CrSBr (device 1). a-e,** Field-dependence of the resistance and magneto-resistance (MR) (bottom panel) as well as its increment (top panel), defined as $\Delta X = X_{+B\rightarrow -B} - X_{-B\rightarrow +B}$, where X states either for the resistance or the MR after sweeping up to different selected magnetic fields at 10 K and $\alpha = \beta = 0°$, being the magnetic field applied in plane along the easy (intermediate) magnetic axis of the top (bottom) CrSBr monolayer. **f,** $\Delta R$ 2D plot. The magnetic sweep protocol is as follows: after a first saturation at negative fields, we perform the sequence ZF $\rightarrow B_{max} \rightarrow -B_{max} \rightarrow$ ZF, increasing in every cycle the maximum field in 20 mT step. Sweeping up (down) trace is depicted in red (blue) in **a-d**. Red/blue arrows indicate the sweeping direction of the magnetic field. MR is defined as MR (%) = 100·[R(B) – R(0)]/R(0), being R(0) the resistance obtained at zero field in the symmetric case.

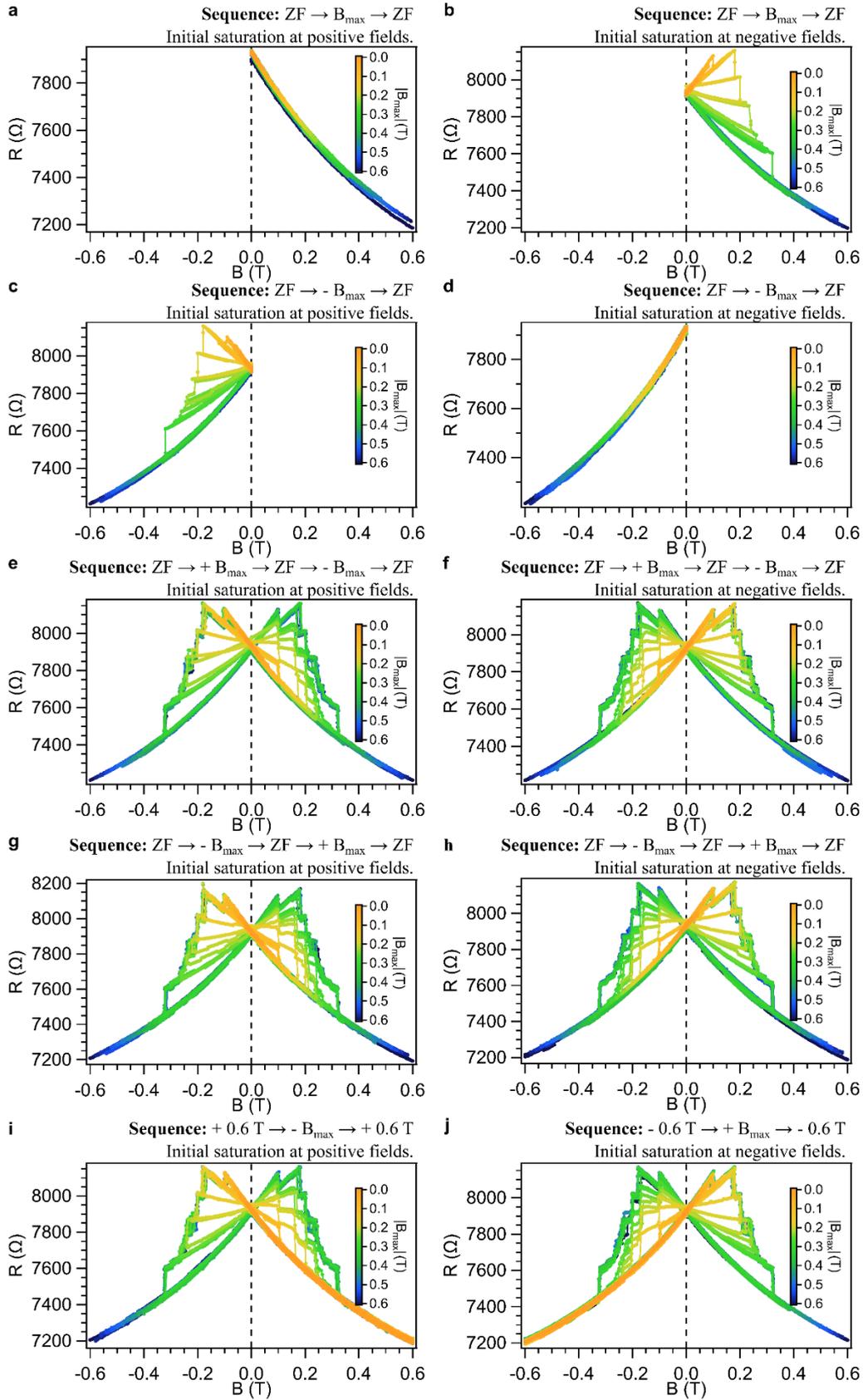

**Supplementary Figure 8.- Multistep spin switching with magnetic memory in orthogonally-twisted CrSBr under different magnetic field sweep protocols.** T = 10 K and $\alpha = \beta = 0\,°$ (device 1). **a,b,** Sequence ZF → +$B_{max}$ → ZF. **c,d,** Sequence ZF → -$B_{max}$ → ZF. **e,f,** Sequence ZF → +$B_{max}$ → ZF → -$B_{max}$ → ZF. **g,h,** Sequence ZF → -$B_{max}$ → ZF → +$B_{max}$ → ZF. **i,** Sequence +0.6 T → -$B_{max}$ → +0.6 T. **j** Sequence -0.6 T → +$B_{max}$ → -0.6 T. Panels **a**, **c**, **e**, **g** and **i** (**b**, **d**, **f**, **h** and **j**) correspond to an initial saturation at positive (negative) magnetic fields. In every field sweep, $B_{max}$ is incremented in steps of 20 mT.

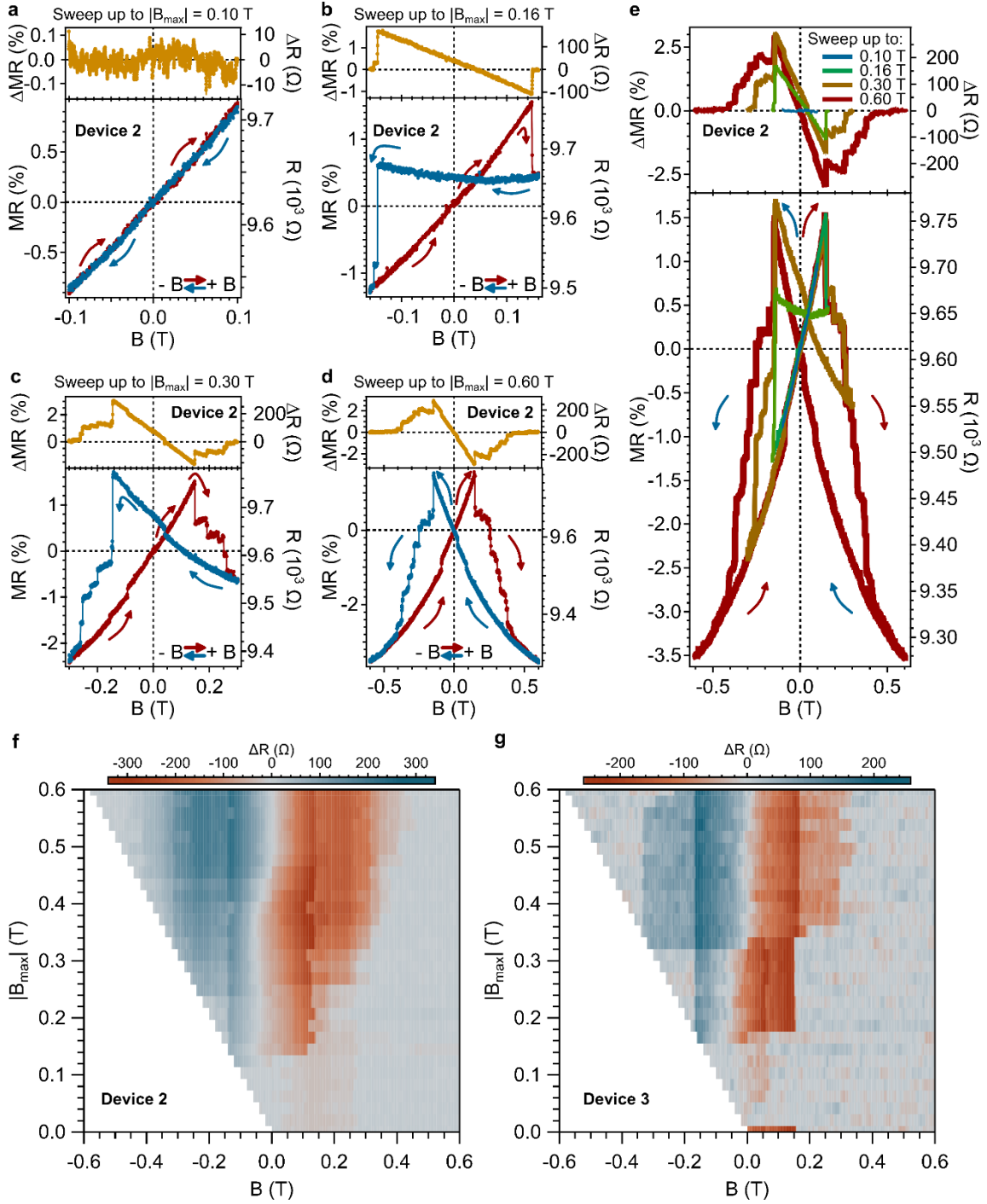

**Supplementary Figure 9.- Hysteresis opening in orthogonally-twisted bilayer CrSBr based on few-layers graphene vertical van der Waals heterostructures. a-e,** Field-dependence of the resistance and magneto-resistance (MR) (bottom panel) as well as its increment (top panel), defined as $\Delta X = X_{+B\rightarrow -B} - X_{-B\rightarrow +B}$, where X states either for the resistance or the MR after sweeping up to different selected magnetic fields at 2 K and $\alpha = \beta = 0$ ° (device 2). **f-g,** $\Delta R$ 2D plot. The magnetic sweep protocol is as follows: for panels **a-e**, after a first saturation at negative fields, we perform the sequence ZF → $B_{max}$ → -$B_{max}$ → ZF, increasing in every cycle the maximum field in 20 mT step. In panel **f-g**, the sequence is +0.6 T → -$B_{max}$ → +0.6 and increasing in every cycle the maximum negative field in 20 mT step, for device 2 (**f**) and device 3 (**g**). Sweeping up (down) trace is depicted in red (blue) in **a-d**. Arrows indicate the sweeping direction of the magnetic field. MR is defined as MR (%) = 100·[R(B) – R(0)]/R(0), being R(0) the resistance obtained at zero field in the symmetric case.

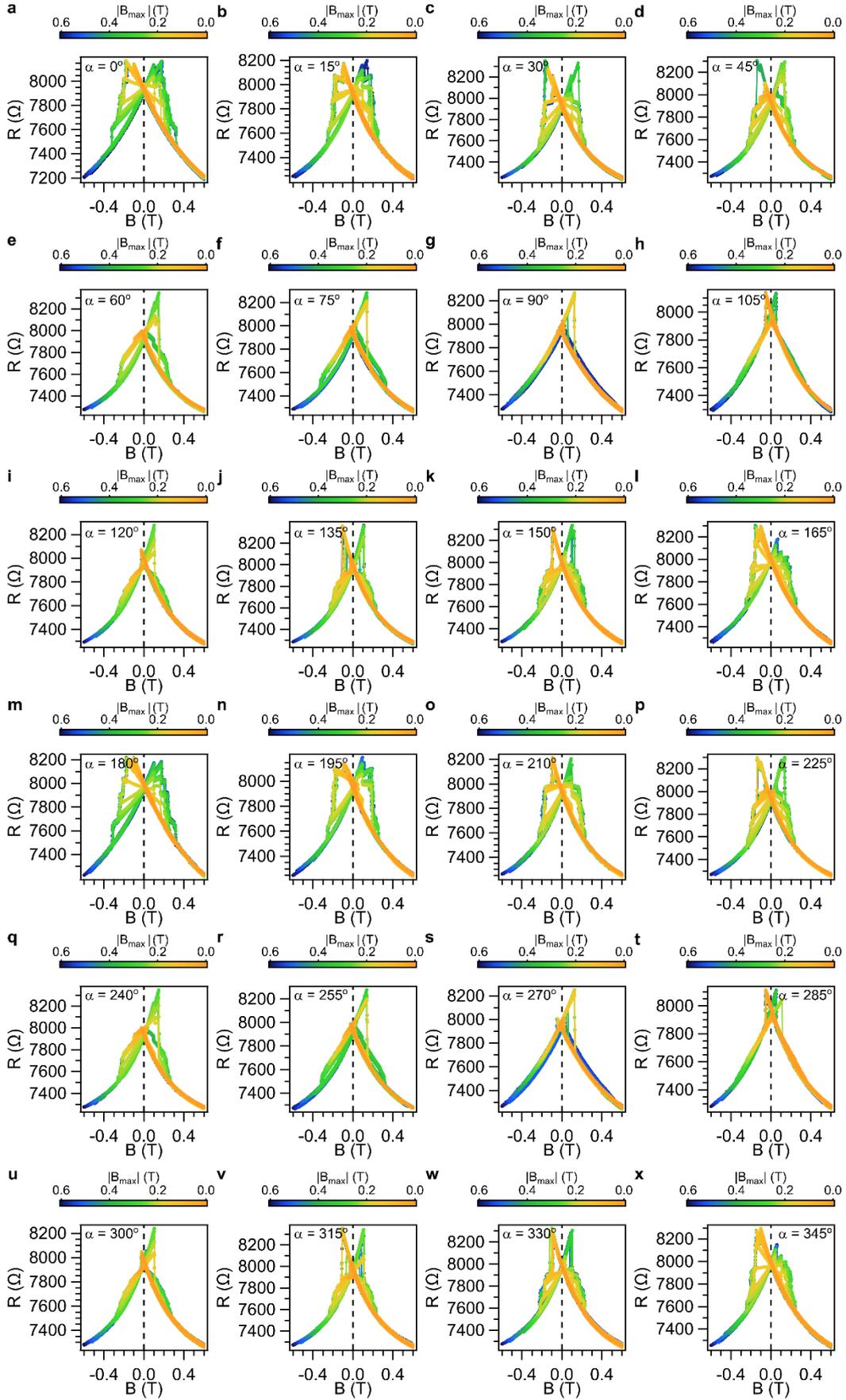

**Supplementary Figure 10.- First-order reversal curves for different in-plane (β = 0 °) magnetic fields at T = 10 K (device 1).** We consider the sequence +0.6 T → -$B_{max}$ → + 0.6 T. In every field sweep, $B_{max}$ is incremented in steps of 20 mT.

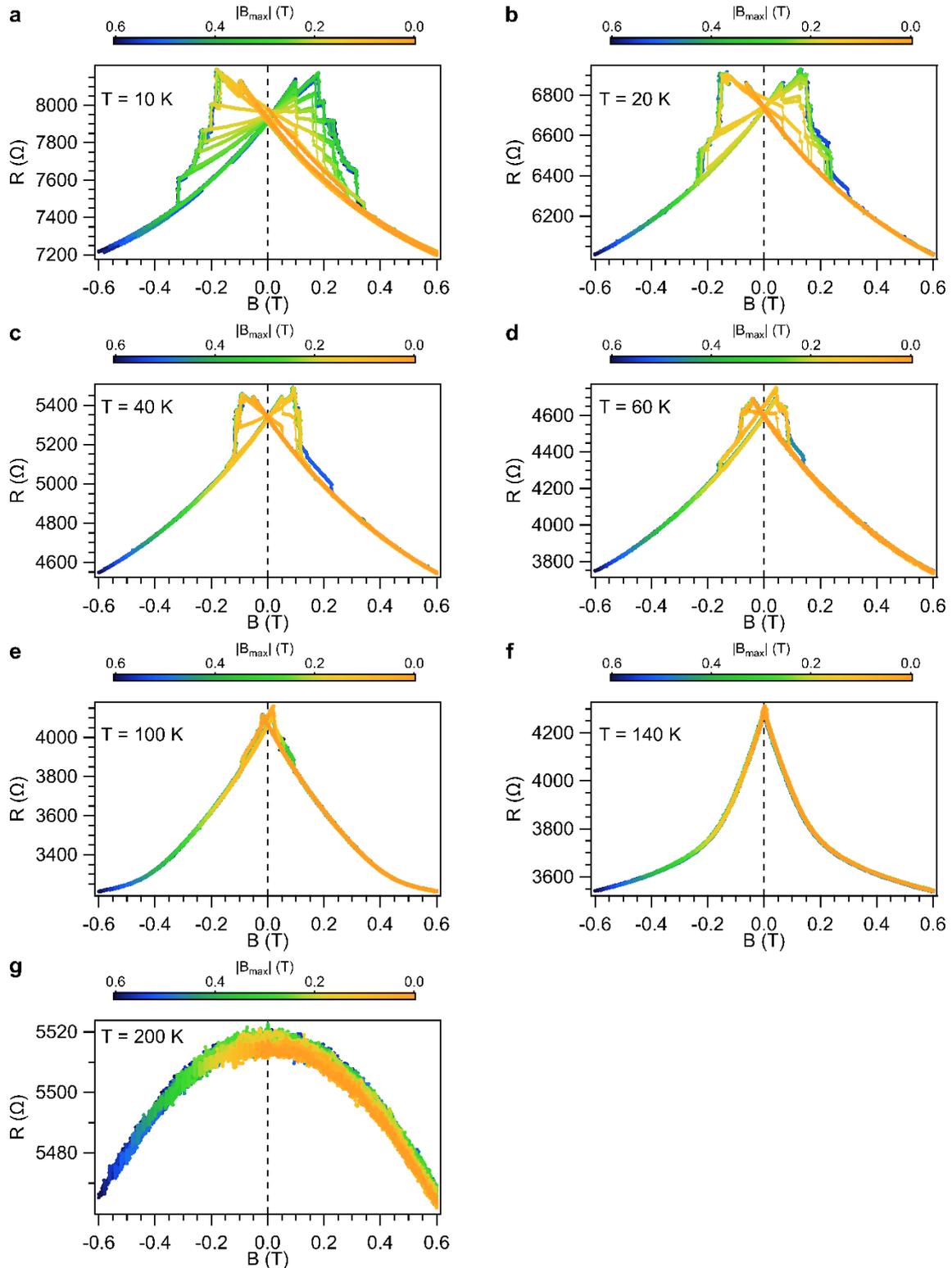

**Supplementary Figure 11.- First-order reversal curves for in-plane (α = β = 0 °) magnetic fields at different temperatures (device 1).** We consider the sequence +0.6 T → -$B_{max}$ → + 0.6 T. In every field sweep, $B_{max}$ is incremented in steps of 20 mT.

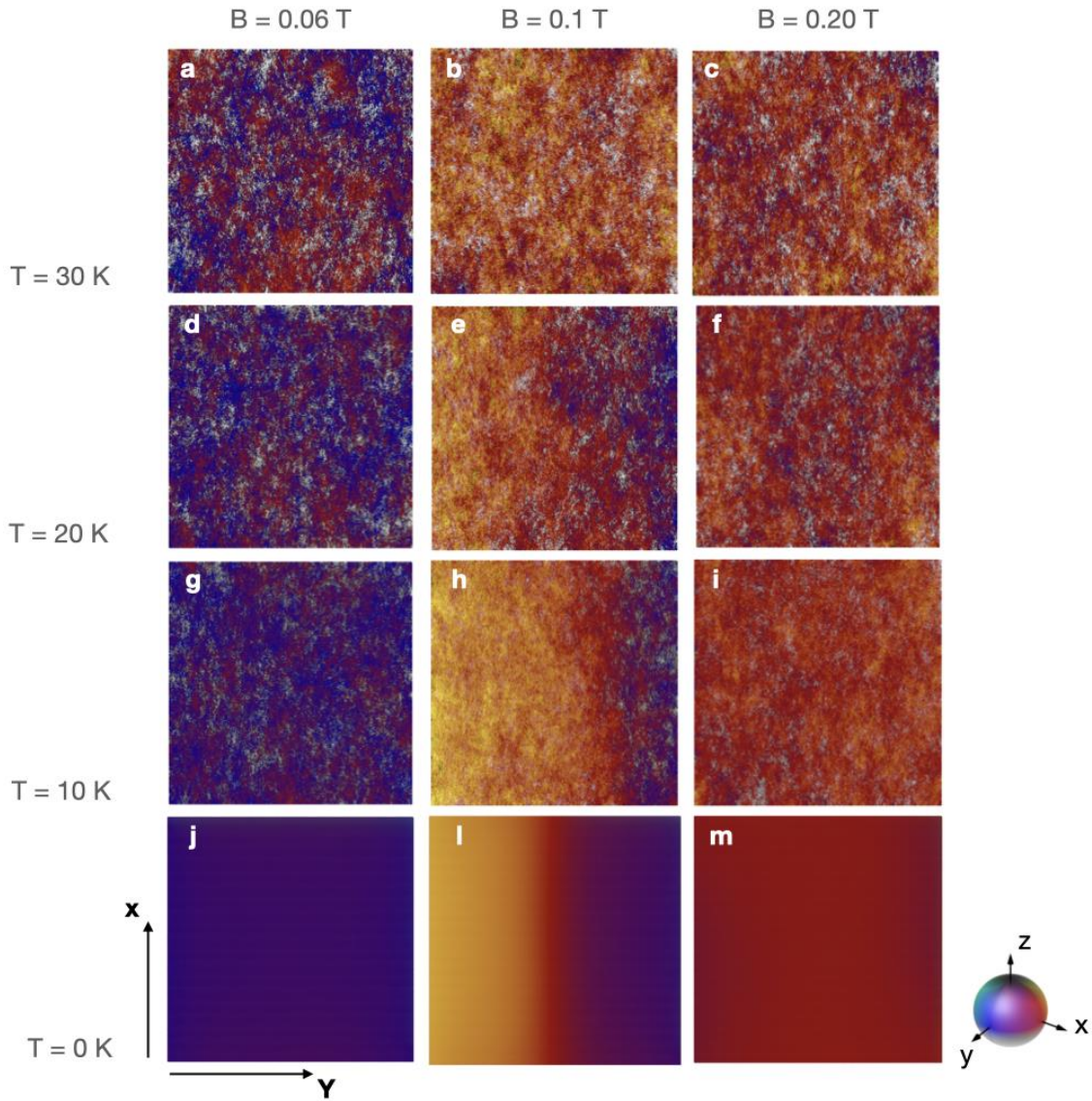

**Supplementary Figure 12.- Spin dynamics simulations for orthogonally-twisted bilayer CrSBr. a-m,** Snapshots of the spin configurations during cooling at different applied fields (0.06 T, 0.10 T, 0.20 T) and temperatures: 30 K (**a-c**), 20 K (**d-f**), 10 K (**g-i**), 0 K (**j-m**). The field is applied following the configuration displayed in the inset of **Figure 4a**. That is, at zero field the easy-axis at both layers are perpendicular to each other due to the device configuration created. As the field is increased, the magnetization of the layer which initially has its easy-axis perpendicular to field (e.g., top layer) rotates to be aligned with the field. The different spin-textures are formed during this process at the corresponding layer not totally oriented with the field. The other layer (e.g., bottom layer) which has its easy-axis already oriented with the external field does not play a substantial role in the phenomena. The spins are oriented accordingly to the magnetic axes displayed in the inset of **Figure 4a**. Each panel measures 100 nm × 100 nm along *y-th* and *x-th* axis with no periodic boundary conditions. Edges have not observed to play any variation on the results.

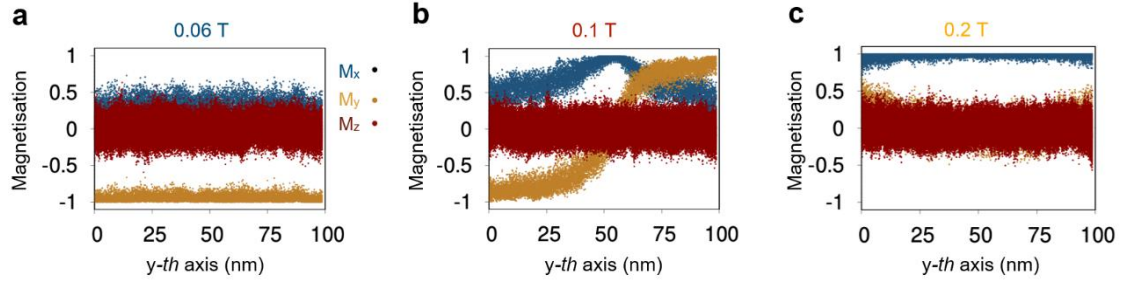

**Supplementary Figure 13.- Field-induced spin-textures in orthogonally-twisted bilayer CrSBr a-c,** Projections of the magnetisation $M_x$, $M_y$ and $M_z$ at 5 K as a function of the position (nm) along the *y*-axis at 0.06 T, 0.1 T and 0.2 T, respectively. See schematic in the inset of **Fig. 4a** for field geometry.

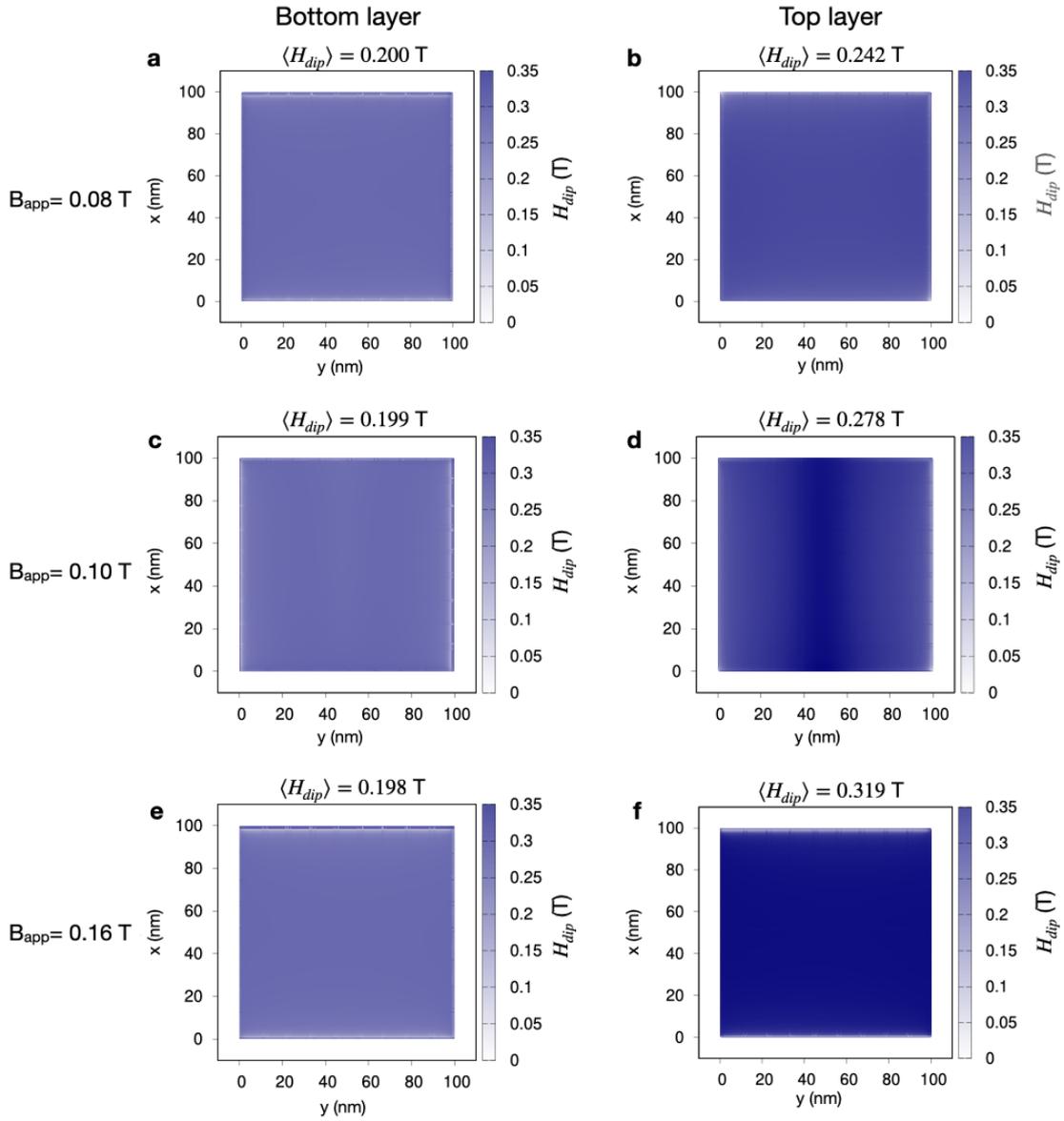

**Supplementary Figure 14.- Dipolar field maps extracted for orthogonally-twisted bilayer CrSBr. a-f,** Dipolar fields ($H_{dip}$) projected over bottom and top layers which have easy-axis parallel and perpendicular, respectively, to the applied field ($B_{app}$). See schematic in the inset of **Fig. 4a** for field geometry. Different magnitudes of $B_{app}$ are applied (0.08 T (**a-b**), 0.10 T (**c-d**) and 0.16 T (**e-f**)), with the corresponding average values of the dipolar fields $\langle H_{dip}\rangle$ induced in the system included at the top of each panel.

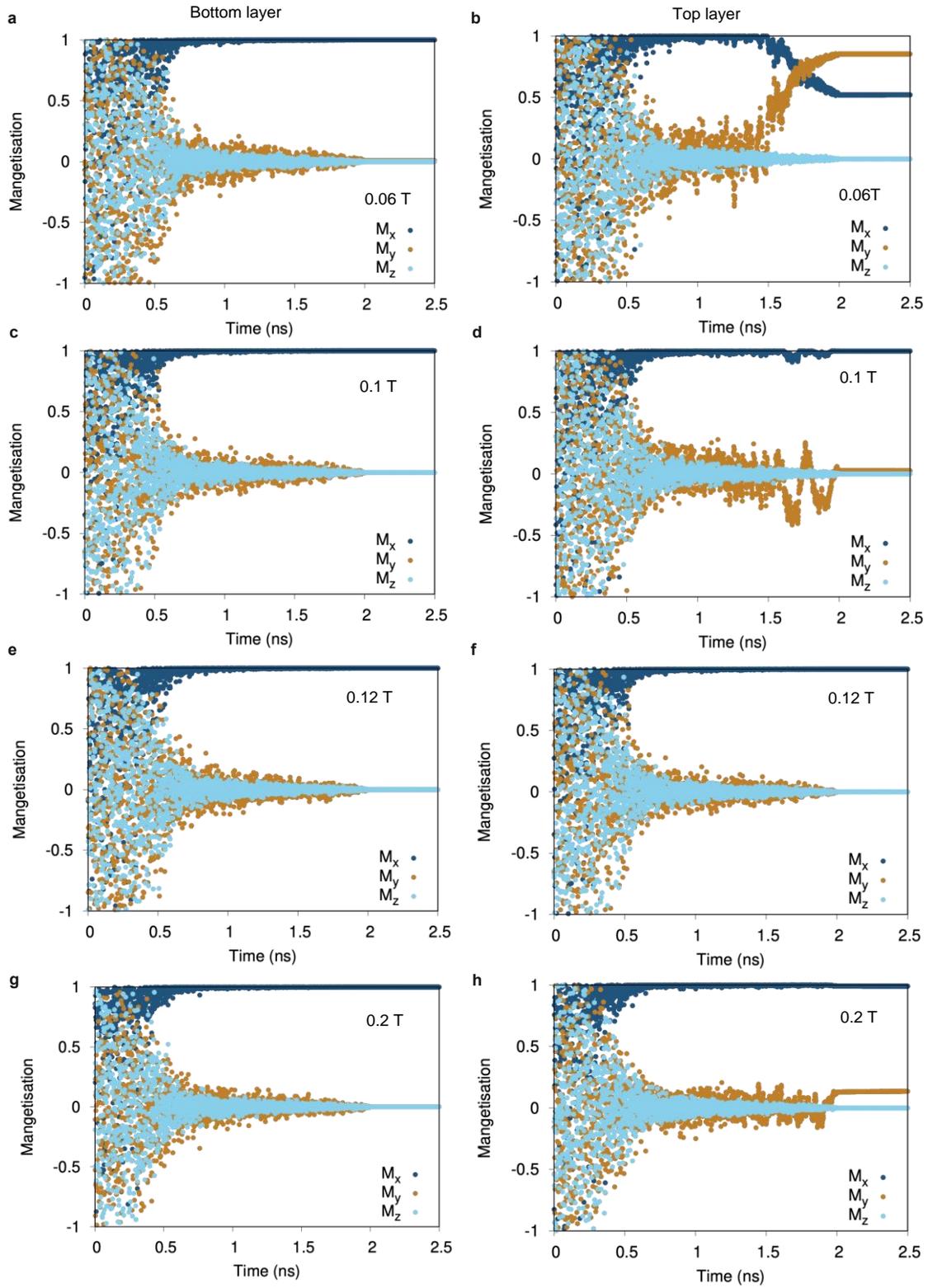

**Supplementary Figure 15.- Time variation of the spins at orthogonally-twisted bilayer CrSBr. a-h,** Projection of the magnetisation along $M_x$, $M_y$ and $M_z$ components for the bottom and top layers as function of time. The magnetic field **B** is applied following the inset of **Figure 4a** and it is parallel ($B_x$) and orthogonal to the easy-axis of the bottom and top layers, respectively. A time interval of 2 ns is recorded for the variations of $M_x$, $M_y$ and $M_z$ under different fields: 0.06 T (**a-b**), 0.1 T (**c-d**), 0.12 T (**e-f**), 0.2 T (**g-h**). The initial spin configurations at 0 ns are randomly assigned at the beginning of the atomistic spin dynamics which generated the large variations of $M_x$, $M_y$ and $M_z$ with time until convergence is achieved.

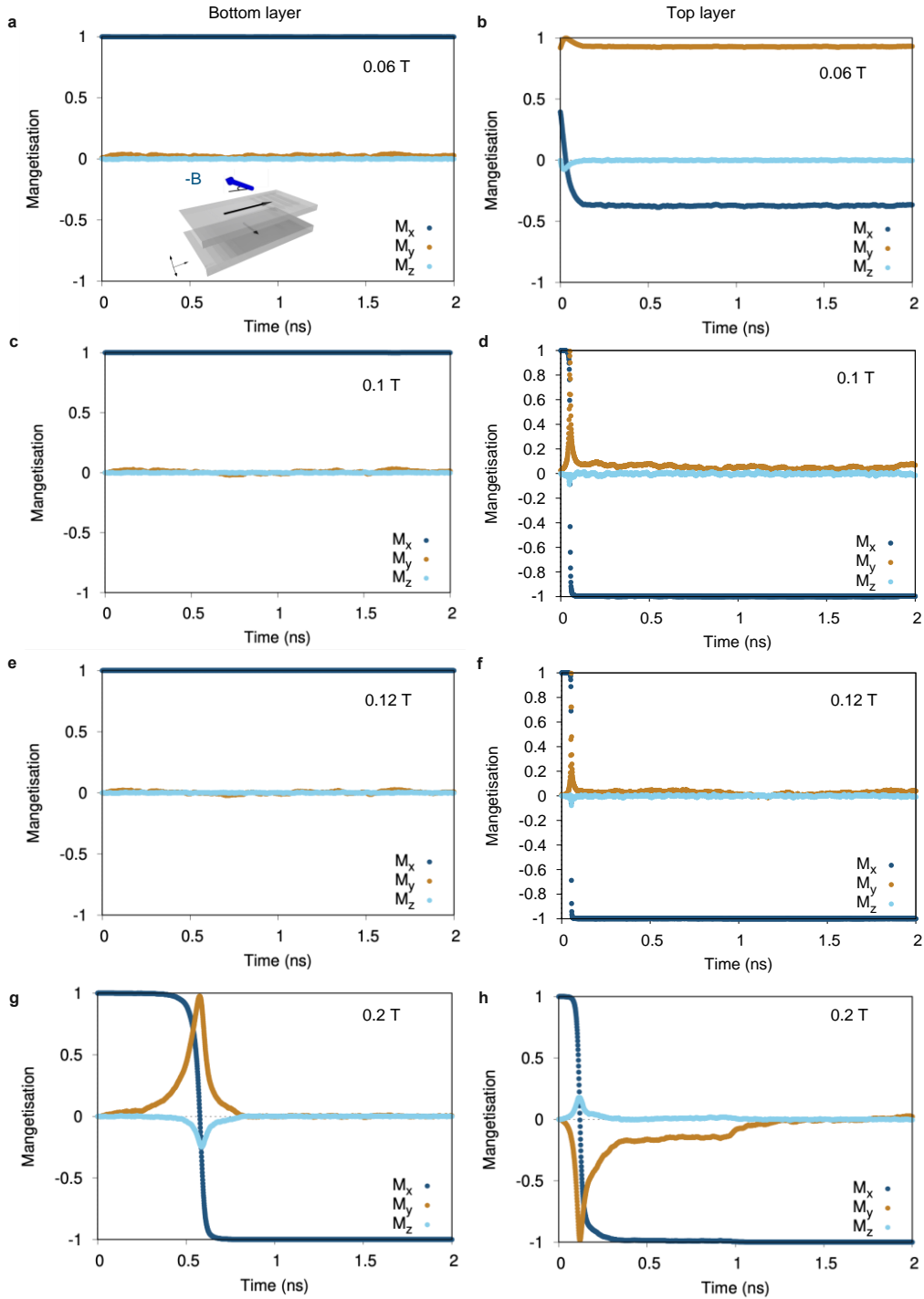

**Supplementary Figure 16.- Time relaxation for spin-flip at orthogonally-twisted bilayer CrSBr at reversal field. a-h,** Projection of the magnetisation along $M_x$, $M_y$ and $M_z$ components for the bottom and top layers as function of time. This figure is similar to **Supplementary Figure 15** but with the magnetic field applied along the opposite direction ($-B_x$). This field is anti-parallel and orthogonal to the easy-axis of the bottom and top layers, respectively. A time interval of 2 ns is recorded for the variations of $M_x$, $M_y$ and $M_z$ under different fields: 0.06 T (**a-b**), 0.1 T (**c-d**), 0.12 T (**e-f**), 0.2 T (**g-h**). The initial spin configurations at 0 ns are those from the simulations with $+B$ showed in **Supplementary Figure 15**. Note that at 0.06 T, the easy-axis of the top layer is tilted between $M_x$ and $M_y$ components since the field is not strong enough to rotate completely the spins. At 0.1 T and 0.12 T a domain wall is initially formed at $+B_x$ situation (**Supplementary Figure 15d,f**) with the average of the spins along the $M_x$ direction. Note that the spins take around 0.2-0.65 ns to change orientation with the applied fields. These finite times induced inhomogeneous magnetic domains and spin textures (**Supplementary Movies S2-S7**) which generate the multi-steps in the magneto-resistance response observed in the devices (**Figure 1**). Since the spins are not totally aligned at intermediate field values at both CrSBr layers, these make the magneto-transport fluctuates abruptly up and down on the resistance. This effect can also be used to approximately quantify how long time the spins take to be fully oriented with the external field.

## Section B – Supplementary Movies 1 – 7

**Supplementary Movie S1:** Multistep magnetization switching with magnetic memory in orthogonally-twisted bilayer CrSBr as shown in **Fig. 3**.

**Supplementary Movie S2**: Movie of the spin-dynamic simulation at field-cooling of 0.06 mT from above 200 K towards 0 K. The magnetic field is applied following the schematic in the inset of **Fig. 4a** with the field parallel to the easy-axis of the bottom layer. The simulation time comprises 2 ns of the cooling process. An additional 1.5 ns simulation-time is undertaking at 0 K to check further the stability The colour scheme follows that in **Suppl. Fig. S11**.

**Supplementary Movie S3:** Similar as **Suppl. Movie S2** at an applied field of 0.10 T. The formation of domain-walls occurred at the top layer as it is still rotating to align with the field.

**Supplementary Movie S4:** Similar as **Suppl. Movie S2** at an applied field of 0.20 T.

**Supplementary Movie S5:** Similar as **Suppl. Movie S2** with a field of 0.06 mT oriented anti-parallel to the easy-axis of the bottom layer (-$B_x$).

**Supplementary Movie S6:** Similar as **Suppl. Movie S5** with a field of $B_x = -0.12$ mT.

**Supplementary Movie S7:** Similar as **Suppl. Movie S4** with a field of $B_x = -0.2$ mT. Note that in general the formation of the domain walls, spin textures, etc. occurred at the top layer (easy-axis perpendicular to the field) despite the direction of the applied field ($\pm B_x$). When the field is reverse (-$B_x$) however, the magnetic structure of the bottom layer (anti-parallel to the field) becomes more inhomogeneous with more fluctuations of the spins.

## Section C – Supplementary Table I

| Intra-monolayer exchange values (meV) | |
|---|---|
| $J_1$ | 7.2920 |
| $J_2$ | 11.0800 |
| $J_3$ | 4.4194 |
| $J_4$ | −0.0032 |
| $J_5$ | −0.0537 |
| $J_6$ | −1.1995 |
| $J_7$ | 0.4293 |
| **Inter-monolayer exchange values (meV)** | |
| $J_{z1}$ | −0.0025 |
| $J_{z2}$ | 0.0025 |

**Supplementary Table I.-** Compendium of the intra and intermonolayer symmetric exchange contributions, $J_i$, used in atomistic spin dynamics.